\documentclass[pra,superscriptaddress,amsmath,amssymb,twocolumn]{revtex4-2}
\usepackage{graphicx}
\usepackage{subfigure}
\usepackage{adjustbox}
\usepackage{bm}
\usepackage{bbm}
\usepackage{color}
\usepackage{braket}
\usepackage{standalone}
\usepackage{multirow}
\usepackage{tikz}
\usepackage{mathrsfs}
\usepackage{dsfont}
\usepackage[colorlinks,bookmarks=true,citecolor=blue,linkcolor=blue,urlcolor=blue]{hyperref}
\usepackage{cleveref}
\usepackage{comment}
\usepackage{mathtools}
\usepackage{soul}

 \Crefname{equation}{Eq.}{Eqs.}
\Crefname{figure}{Fig.}{Figs.}

\begin{document}
\title{Stability of dipolar bosons in a quasiperiodic potential}

\author{Paolo Molignini}
\affiliation{Department of Physics, Stockholm University, AlbaNova University Center, 10691 Stockholm, Sweden}
\author{Barnali Chakrabarti}
\affiliation{Department of Physics, Presidency University, 86/1 College Street, Kolkata 700073, India}
\affiliation{Instituto de Física, Universidade de São Paulo, São Paulo, SP CEP 05508-090, Brazil}
\date{\today}

\begin{abstract}
Quasiperiodic potentials and dipolar interactions each impose long-range order in quantum systems, but their interplay unlocks a rich landscape of unexplored quantum phases. 
In this work, we investigate how dipolar bosonic crystals respond to correlated disorder in the form of quasiperiodic potentials. 
Using exact numerical simulations and a suite of observables -- including order parameters, energy, density distributions, and two-body coherence measures -- we explore one-dimensional dipolar bosons in quasiperiodic lattices at both commensurate and incommensurate fillings. 
Our results reveal a complex competition between superfluid, Mott insulator, density-wave, and crystalline phases, governed by the intricate balance of dipolar interactions, kinetic energy, and disorder strength. 
Crucially, we identify mechanisms that influence dipolar crystals, showing their surprising robustness even in the presence of strong quasiperiodic disorder. 
Strikingly, we challenge previous claims by demonstrating that a \emph{kinetic} crystal phase --expected to precede full crystallization -- does not emerge in the ground state. 
Instead, its traits appear only under moderate disorder, but never fully develop, giving way to a direct transition from a charge density wave to a crystal state. 
These findings provide new insights into the resilience of many-body quantum phases in complex environments and pave the way for engineering exotic quantum states in ultracold atomic systems.
\end{abstract}

\maketitle

\section{Introduction} 

Dipolar quantum systems have emerged as a promising platform for exploring exotic quantum phases driven by long-range interactions~\cite{Lahaye:2009, Gadway:2016, Moses:2017, Norcia:2021natphys, Chomaz:2023, Defenu:2023}.
Among these, dipolar ``crystal states" -- characterized by highly ordered structures stabilized by dominant dipolar repulsions -- have garnered significant attention~\cite{Deuretzbacher:2010, Zoellner:2011-PRA, Zoellner:2011-PRL, Fischer:2015, Chatterjee:2018, Chatterjee:2019, Chatterjee:2020}. 
These states are particularly intriguing in low-dimensional setups, where the interplay between geometry and dipolar interactions leads to the formation of robust crystal-like configurations in various arrangements, such as zigzag chains~\cite{Astrakharchik:2008-PRA77, Astrakharchik:2008-PRA78}, rings~\cite{Zoellner:2011-PRA}, and triple wells~\cite{Chatterjee:2013}.
The study of dipolar crystal states is not only theoretically fascinating but also pivotal for understanding the emergence of novel quantum orders that lie beyond traditional condensed matter paradigms. 
Furthermore, the remarkable stability of these crystal states in one-dimensional systems positions them as ideal candidates for investigating fundamental properties of long-range ordered phases, and also opens up avenues for potential applications in quantum technology~\cite{DeMille:2002, Rabl:2007, Weimer:2012, Islam:2013}. 

Ultracold atoms provide an exceptionally versatile and robust platform for realizing and tuning dipolar interactions. 
Atomic species such as erbium $^{167}$Er~\cite{Aikawa:2012}, dysprosium $^{161}$Dy~\cite{Lu:2011}, and chromium $^{53}$C~\cite{Griesmaier:2005}, as well as polar molecules like potassium-rubidium $^{40}$K$^{87}$Rb~\cite{Ni:2008}, sodium-lithium $^{23}$Na$^{6}$Li~\cite{Rvachov:2017}, and sodium-potassium $^{23}$Na$^{40}$K~\cite{Park:2015}, exhibit strong dipole-dipole interactions that can be precisely controlled in laboratory conditions. 
The unparalleled experimental control over interaction strengths, dimensionality, and trapping geometries in ultracold setups enables the realization of regimes where dipolar interactions dominate over short-range forces.
This tunability can lead to the emergence of novel quantum phases, including $p$-wave superfluids~\cite{Bruun:2008, Cooper:2009}, density-wave phases~\cite{DallaTorre:2006, Maluckov:2012}, Haldane insulators~\cite{Deng:2011}, Luttinger-liquid-like behaviors~\cite{Citro:2007, Citro:2008}, checkerboard orderings~\cite{Menotti:2007}, Mott solids~\cite{Capogrosso-Sansone:2010}, and supersolid phases with coexistent density wave and superfluidity~\cite{Pollet:2010, Boettcher:2019, Tanzi:2019,  Chomaz:2019, Guo:2019, Natale:2019, Tanzi:2019-2, Tanzi:2021, Norcia:2021, Sohmen:2021, Sanchez-Baena:2023, Recati:2023}, thus establishing ultracold atoms as a cornerstone for exploring dipolar many-body physics.

The intersection of dipolar interactions with disorder opens exciting opportunities to investigate competing ordering mechanisms. 
Controlled disorder has revealed phenomena such as Bose glass phases~\cite{Giamarchi:1988, Fisher:1989, Lye:2005, Palencia:2019}, many-body localization~\cite{Bloch:2019, Oganesyan:2020}, unconventional Mott lobes~\cite{Palencia:2019, Yao:2020, Chakrabarti:2024}, and Anderson localization~\cite{Lugan:2007, Lugan:2007-2, Palencia:2010, Billy:2008, Lugan:2011}. 
Meanwhile, \emph{quasiperiodic} potentials offer a distinct form of correlated disorder, characterized by long-range structural order without translational symmetry.
The rapid advancement in trapping cold atoms has made possible the realization of quasiperiodicity and thus quasicrystalline structures with extreme precision~\cite{Fallani:2007, Bordia:2017, Lueschen:2017, Nakajima:2021, Gautier:2021, Yao:2024}. 
Quasicrystals can be engineered with ultracold atoms using controlled interactions~\cite{Roati:2008, Roux:2008, Billy:2008, Palencia:2010, Modugno:2010, Derrico:2014, Palencia:2020, Derrico:2021, Yao:2024} or photon-mediated interactions~\cite{Hou:2018, Mivehvar:2019}, and have been theoretically studied both in one and two dimensions~\cite{Roux:2008,  Palencia:2020, Gautier:2021, Zhu:2023}.
Among the different setups to realize the localization effect, one-dimensional (1D) bichromatic quasiperiodic optical lattices -- as used in the experiment of Roati {\it et al.}~\cite{Roati:2008} -- have gained particular interest.
Quasiperiodic systems exhibit unique properties like topological order~\cite{Lang:2012, Dareau:2017, Baboux:2017}, fractality~\cite{Yuan:2000, Tanese:2014} and unusual localization behaviors~\cite{Roati:2008, Molignini:2025-quasicryst}.
Moreover, the combination of quasiperiodicity with dipolar interactions allows for the study of phases where long-range dipolar ordering competes with quasiperiodic structural constraints~\cite{Senechal:1995}. 
The intricate interplay between these mechanisms promises to uncover novel phases, including compressible insulators and distorted Mott states, with implications for both fundamental physics and applications.

While initial studies have begun to explore the influence of dipolar interactions on quasicrystals, many open questions remain. 
For instance, can strong dipolar interactions stabilize or modify quasiperiodic ordering in novel ways? 
Conversely, how does quasiperiodicity affect the stability of dipolar crystal states? 
While early studies suggest dipolar interactions can induce localization phenomena that both enhance and compete with quasiperiodicity~\cite{Molignini:2025-quasicryst}, the interplay between these competing forms of long-range order is largely uncharted territory. 
Understanding how disorder, incommensuration, and quasiperiodicity interact with dipolar crystal phases could provide profound insights into the emergence of new states, such as charge-density waves, fragmented Mott insulators, and nontrivial superfluid phases.

This paper aims to study the pathway to dipolar crystal states and their stability under quasiperiodic potentials. 
We employ the MultiConfigurational Time-Dependent Hartree method~\cite{Streltsov:2006, Streltsov:2007, Alon:2007, Alon:2008, Lode:2016, Fasshauer:2016, Lin:2020, Lode:2020, Molignini:2025-SciPost, MCTDHX} to solve the Schrödinger equation for dipolar bosons in one-dimensional quasiperiodic lattices. 
We investigate how the competition between dipolar interactions, kinetic energy, and one-body potentials gives rise to a variety of density-modulated phases. 
The focus is on identifying the mechanisms that stabilize or destabilize crystal states in both commensurate and incommensurate fillings, thereby expanding our understanding of long-range interacting systems in disordered environments.

Our results show that commensurate fillings support superfluid, Mott insulator, and crystal states, while incommensurate fillings also give rise to charge density waves. 
We demonstrate how disorder and incommensuration interact with dipolar interactions to influence localization in intermediate phases. 
Our findings reveal a complex interplay between long-range interactions and quasiperiodic disorder, leading to rich pathways and stability mechanisms in dipolar bosonic crystals. 
Correlated disorder significantly modifies the competition between superfluid, Mott insulator, and density-modulated states, stabilizing unexpected localization patterns. 
Notably, quasiperiodicity enhances superfluidity at weak interactions but has minimal impact on crystal state robustness, which persists even in strongly disordered environments. 
Moreover, contrary to previous predictions~\cite{Chatterjee:2020}, we find no evidence of a ``kinetic crystal phase" in the ground state at incommensurate filling, as the crystal state emerges directly via interaction-driven dimerization of the charge density wave. 
These results provide new insights into long-range order in complex quantum environments and offer avenues for engineering many-body states in ultracold atomic experiments.

The rest of this paper is structured as follows.
In section \ref{sec:system}, we present the theoretical framework behind the system we engineer and study.
In section \ref{sec:methods}, we give a brief review of the method we employ to investigate the system numerically exactly. 
In this part, we also define the observables used to map out the different features of the system.
In sections \ref{sec:results-comm} and \ref{sec:results-incomm} we present our results for two setups: one with commensurate particle filling of $\nu=2$ (two particles per site) and another with an incommensurate particle filling of $\nu=8/5$.
Section \ref{sec:conclusions} concludes our presentation with a summary of the main findings and an outlook on potential future research directions.

\section{System and protocol}
\label{sec:system}
We study the ground state of $N$ interacting bosons of mass $m$ in a one-dimensional quasiperiodic lattice.
The system is described by the time-dependent many-body Schr\"odinger equation 
\begin{equation}
\hat{H} \Psi = i \hbar \frac{\partial \Psi}{\partial t}
\label{eq:TISE}
\end{equation}
with the many-body Hamiltonian
\begin{equation} 
\hat{H}(x_1,x_2, \dots x_N)= \sum_{i=1}^{N} \hat{h}(x_i) + \sum_{i<j=1}^{N}\hat{W}(x_i - x_j).
\label{eq:Hamiltonian}
\end{equation}
The one-body part of the Hamiltonian is $\hat{h}(x) = \hat{T}(x) + \hat{V}_{OL}(x)$, where $\hat{T}(x) = -\frac{\hbar^2}{2m} \frac{\partial^2}{\partial x^2}$ is the kinetic energy operator and $\hat{V}_{OL}(x) = V_p \sin^2(k_p x) + V_d \sin^2(k_d x + \phi)$ is an external potential characterized by two interfering standing waves that generate a quasiperiodic lattice.

The quasiperiodic potential consists of a primary lattice of depth $V_p$ and wave vector $k_p$ perturbed by a detuning lattice of depth $V_d$, wave vector $k_d$, and phase shift $\phi$.
When the ratio of the two wave vectors $k_p/k_d$ is chosen to be incommensurate, the external potential generates a regular but non-repeating structure that implements correlated disorder on top of the periodic primary lattice.
Unless otherwise stated, in our study we will set $\phi=0$ for simplicity.
The introduction of correlated disorder has two main effects for the small-scale systems we consider.
First of all, it lifts the degeneracy of all the potential minima by creating a global minimum (at $x=0$ for $\phi=0$).
Moreover, it shifts the occurrence of the other (now) local minima away from the periodic case $V_d=0.0$, breaking its translational invariance.

The Hamiltonian in Eq.~\eqref{eq:Hamiltonian} can be written in dimensionless units obtained by dividing the dimensionful Hamiltonian by $\frac{\hbar^2}{m\bar{L}^2}$, with $\bar{L}$ an arbitrary length scale.
For the sake of our calculations, we will set $\bar{L}$ to be the period of the primary optical lattice.
Unless otherwise stated, we will keep the wave vectors fixed at $k_p=1.0$ and $k_d=1.19721$.
We will instead probe various values of the potential depths $V_p \in [6 E_r, 16 E_r]$ and $V_d \in [0 E_r, 6 E_r]$, where $E_r = \frac{\hbar^2 k_p^2}{2 m}$ is the recoil energy of the primary lattice.
Depending on the number of particles considered, we restrict the system geometry to accommodate $S=3$ or $S=5$ sites in the primary optical lattice (defined as the spatial extent between two maxima in the sinusoidal function) adding hard-wall boundary conditions at each end.
The two cases will be used to probe the physics of both commensurate particle filling ($N=6$ particles in $S=3$ sites) and incommensurate filling ($N=8$ particles in $S=5$ sites).

The two-body interactions $\hat{W}(x_i - x_j)$ appearing in Eq.~\eqref{eq:Hamiltonian} are modelled with long-range dipole-dipole interactions (DDI) $W_D(x_i,x_j) = \frac{g_d}{|x_i-x_j'|^3 + \alpha}$.
Here, $g_d$ controls the strength of the DDI and it is positive (repulsive interactions) for all our calculations.
To probe the pathway from superfluid state (at low $g_d$) to crystal state (at high $g_d$), we will study particles with interaction strengths over five orders of magnitude in the range $g_d \in [0.002 E_r, 20 E_r]$.
The parameter $\alpha$ denotes a renormalization factor that arises when DDI are confined to a one-dimensional geometry~\cite{Deuretzbacher:2010, Sinha:2007, Cai:2010}.
We will keep it constant at $\alpha=0.05 \bar{L}^3$ for all our computations.

The Hamiltonian defined above is an accurate depiction of the energy landscape engineered in state-of-the-art experiments with magnetic atoms or dipolar molecules loaded in optical lattices.
At low temperature, the interactions between neutral atoms characterized by the $s$-wave scattering length are changed by tuning the external magnetic field in the vicinity of Feshbach resonance~\cite{Tiesinga:1992}. 
For the ensemble of atoms with large magnetic moment, the dipole-dipole interactions can also be tuned experimentally~\cite{Tang:2018,Li:2021,Giovanazzi:2022}.
We will solve the Hamiltonian above numerically exactly to extract the ground state properties of the system and map out relevant observables such as energy, real-space density, and correlations.
These observables will reveal the fate of the many-body state subject to both long-range interactions and quasiperiodicity.


\section{Methods}
\label{sec:methods}
To study the ground-state properties of dipolar bosons with correlated quasiperiodic disorder we employ the MultiConfigurational Time-Dependent Hartree method for indistinguishable particles (MCTDH)~\cite{Streltsov:2006, Streltsov:2007, Alon:2007, Alon:2008,Lode:2016,Fasshauer:2016}.
The MCTDH method addresses the many-body Schrödinger equation by representing the many-body wave function as an adaptive, time-dependent superposition of permanents, which are built from \( M \) single-particle wave functions known as orbitals. 
This approach involves optimizing both the coefficients and the basis functions in the superposition over time, enabling the calculation of ground-state properties through imaginary relaxation or the simulation of full-time dynamics using real-time propagation.
In this work, we focus exclusively on imaginary time propagation to relax the system to its ground state.
For our numerics, we employ the MCTDH-X software~\cite{Lin:2020, Lode:2020, Molignini:2025-SciPost, MCTDHX}, which is the most advanced algorithmic implementation of the MCTDH approach.
Further details on this method are contained in Appendix~\ref{app:MCTDHX} .

Utilizing several orbitals in the many-body ansatz allows us to capture fragmented many-body states -- like crystal states -- that cannot be described by mean-field or mean-field-adjacent approximations like extended Gross-Pitaevskii equations.
The convergence of the method is guaranteed in the limit $M \to \infty$. 
However, in practice a much smaller set of orbitals is needed to obtain \emph{numerically exact} results, i.e. results that do not change by including more orbitals in the MCTDH expansion.
In all our calculations, we employ $M=12$ orbitals and find that our results do not qualitatively change upon adding more orbitals.
In particular, the ground state energy is converged up to 12 decimal digits.

To study the ground state of the system, we compute various observables from the many-body state $\left| \Psi \right>$.
The most immediate observable is given by the many-body energy, which can be separated in kinetic, potential, and interacting energy:
\begin{align}
E_{\mathrm{kin}} &= \left< \Psi \right| \sum_i \hat{T}(x_i) \left| \Psi \right> \\
E_{\mathrm{pot}} &= \left< \Psi \right| \sum_i \hat{V}_{OL}(x_i) \left| \Psi \right> \\
E_{\mathrm{int}} &= \left< \Psi \right| \sum_{i,j} \hat{W}(x_i - x_j) \left| \Psi \right>.
\end{align}
The breakdown in the different energy components allows us to deduce what is the dominant energy scale in each phase.

To examine the spatial arrangement of the bosons in each phase, we compute the one-body density as
\begin{align}
\rho(x) &= \left< \Psi \right| \hat{\Psi}^{\dagger}(x) \hat{\Psi}(x) \left| \Psi \right>,
\end{align}
where $\Psi^{(\dagger)}(x)$ creates (annihilates) a boson at site $x$.
Additionally, to detect the effect of quasiperiodic disorder on different phases, it is useful to obtain a measure of population imbalance between odd and even lattice sites in the primary optical lattice.
This is defined as
\begin{equation}
\mathcal{I} = \sum_{x} \left( \rho_{\mathrm{odd}}(x) - \rho_{\mathrm{even}}(x) \right),
\end{equation}
where $\rho_{\mathrm{odd}/\mathrm{even}}$ is the particle density at odd/even sites.
This measure has been previously shown to be sensitive to changes in the nature of the crystal state~\cite{Chatterjee:2020}.

To measure the degree of coherence and many-body correlation, we calculate 
the reduced one-body (1-RDM) and two-body (2-RDM) density matrices:
\begin{align}
\rho^{(1)}(x,x') &= \left< \Psi \right| \hat{\Psi}^{\dagger}(x) \hat{\Psi}(x') \left| \Psi \right> \\
\rho^{(2)}(x,x') &= \left< \Psi \right| \hat{\Psi}^{\dagger}(x) \hat{\Psi}^{\dagger}(x') \hat{\Psi}(x') \hat{\Psi}(x) \left| \Psi \right>.
\end{align}
From the 1-RDM, one can also extract information about the orbital occupations in the ground state.
This is described in terms of the natural orbitals $\phi_i^{(\mathrm{NO})}$ and their occupations $\rho_i$, which correspond to the eigenfunctions and eigenvalues of $\rho^{(1)}(x,x')$, respectively, as given by
\begin{equation}
\rho^{(1)}(\mathbf{x},\mathbf{x}') = \sum_i \rho_i \phi^{(\mathrm{NO}),*}_i(\mathbf{x}')\phi^{(\mathrm{NO})}_i(\mathbf{x}).
\label{eq:RDM1}
\end{equation}
Besides providing information about orbital occupation and convergence, the eigenvalues of the 1-RDM can be used to construct an order parameter~\cite{Chatterjee:2018, Chatterjee:2020}
\begin{equation}
\Delta = \sum_i \left( \frac{\rho_i}{N} \right)^2
\label{eq:order-par}
\end{equation}
that quantifies the loss of coherence and thus maps out the transition from superfluid ($\Delta \to 1$) to crystal state ($\Delta \to \frac{1}{N}$).
 

\section{Results in periodic and quasiperiodic lattice}

We now present an exhaustive study of the ground-state properties of dipolar-interacting bosons in periodic and quasiperiodic lattices.
We begin by detailing our numerical results for the commensurate filling case of $N=6$ bosons in $S=3$ sites in Sec.~\ref{sec:results-comm}, and then focus on the incommensurate filling case of $N=8$ bosons in $S=5$ sites in Sec.~\ref{sec:results-incomm}.

\subsection{Commensurate filling}
\label{sec:results-comm}

To contextualize our results in quasiperiodic lattices, we first summarize the known physical behavior of repulsive dipolar bosons in periodic lattices. 
For dipolar atoms in a periodic lattice we expect three possible states: compressible superfluid (SF), incompressible Mott-insulator (MI), and compressible crystal state (CS).
A SF is dominated by kinetic energy over interactions, and with imposed hard-wall boundary condition exhibits a single orbital contributing to the vast majority of a strongly localized wave function  (typically resulting in $\Delta \gtrsim 0.9$).
A MI exhibits a more or less equally distributed density over multiple orbitals and sites.
For our current setup, this should result in doubly-populated peaks at all three sites, with a marked drop in the order parameters to values $\Delta \sim \frac{1}{S} = \frac{1}{3}$.
In the CS, the double occupation of each peak is further lifted by the strong interparticle repulsions and six separate peaks emerge. 
Importantly, the center of mass of these peaks is shifted away from the potential minima due to the dipolar repulsions.
We also remark that this phase is only observed in long-range interacting systems.
Due to its localization properties, the CS must be described by exactly $M=N=6$ macroscopically populated orbitals, resulting in an order parameter $\Delta \sim \frac{1}{M} = \frac{1}{6}$. 
Moreover, since the interactions represent the dominant energy scale in the crystal state, we expect it to be relatively unaffected by the potential depth.

\begin{figure}[h!]
\centering
\includegraphics[width=1.0\columnwidth]{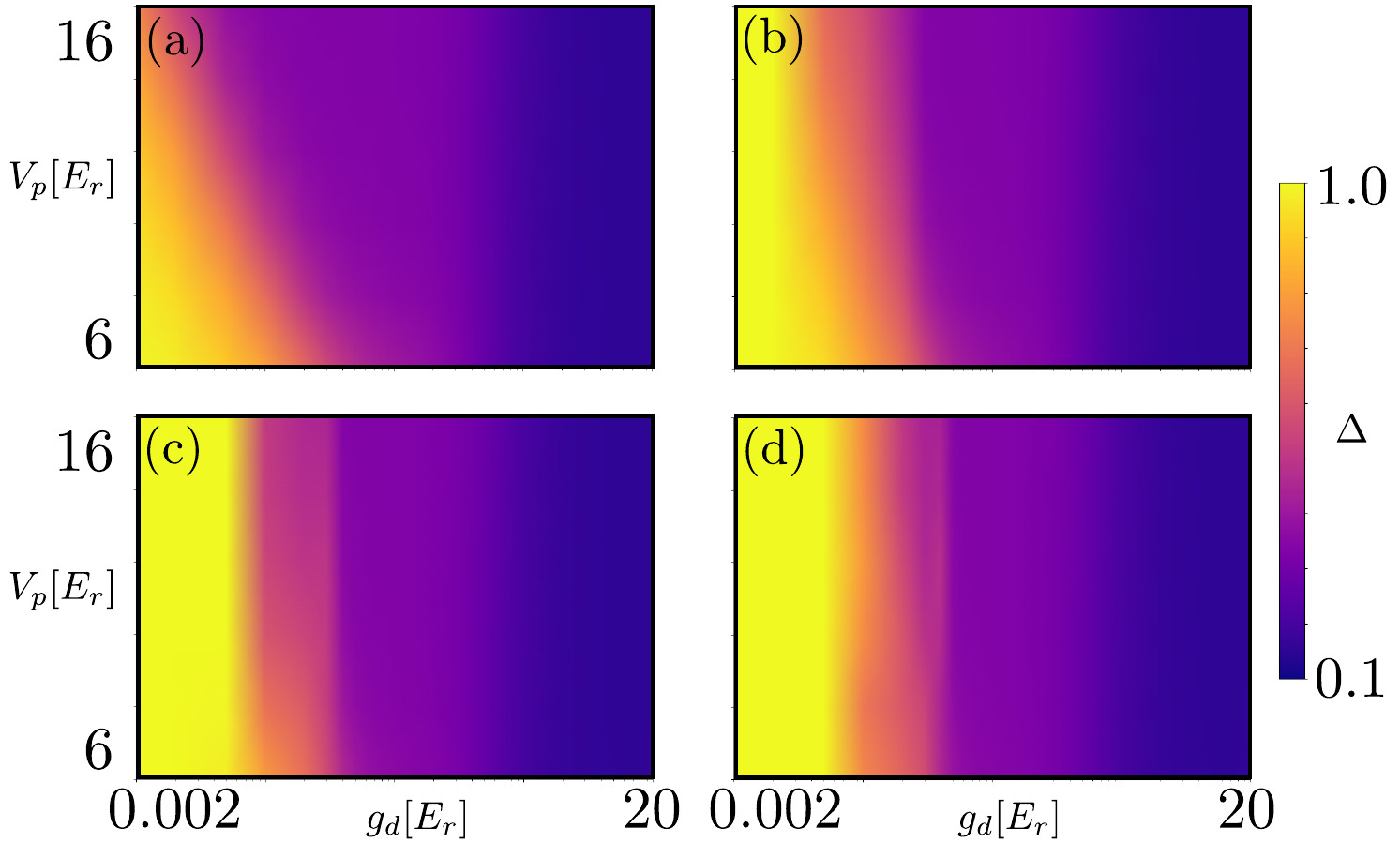}
\caption{Order parameter diagrams for the commensurate filling case $N=6$, $S=3$. 
The plots showcase the order parameter $\Delta$ as a function of the primary lattice depth $V_p$ and the interaction strength $g_d$.
The four panels show results for four different disorder potential depths:
(a) $V_d=0.0 E_r$,
(b) $V_d=2.0 E_r$,
(c) $V_d=4.0 E_r$,
(d) $V_d=6.0 E_r$.
}
\label{fig:PD-commensurate}
\end{figure}

\subsubsection{Order parameter}
We now present our numerical results.
As a primary measure for the emergence of the different states in the commensurate setup, we utilize the order parameter $\Delta$ (cf. Eq.~\ref{eq:order-par}).
Fig.~\ref{fig:PD-commensurate} summarizes our main results in the form of phase diagrams in the parameter space spanned by the primary lattice depth $V_p$ and interaction strength $g_d$, obtained at different values of the detuning lattice depth $V_d=0, 2, 4$, and $6 E_r$.
We cover the entire range of weak to deep lattices $V_p \in \left[6.0 E_r, 16.0 E_r\right]$ and  interaction strengths over four orders of magnitude, i.e. $g_d \in \left[0.002 E_r, 20 E_r\right]$.

As we can see in Fig.~\ref{fig:PD-commensurate}(a), our numerics perfectly recreates the expected behavior of the periodic system.
The SF phase (yellow region) is confined to small barrier heights \(V_p\) and low interaction strengths \(g_d\). 
As \(V_p\) increases, the MI phase gradually emerges (orange, red, and purple regions). 
As known from the literature~\cite{Lin:2019}, this phase transition is of second-order and characterized by a progressive fragmentation of the superfluid.
With a further increase in \(g_d\), the insulating phase persists across the entire range of lattice depths. 
This indicates that, within the periodic limit, independent control of lattice depth and interaction strength allows the realization of Mott phases under two conditions: either when the interaction strength is sub-dominant and the lattice potential is dominant, or when the interaction strength is dominant and the lattice potential is sub-dominant. 
This behavior contrasts with the Bose-Hubbard model, where the transition from the superfluid (SF) to the MI phase is governed solely by the ratio \(\frac{U}{J}\), where \(U\) represents the interaction strength and \(J\) the hopping term, regardless of how this ratio is achieved. 
As \(g_d\) increases further, crystalline states (dark blue region) emerge. 
Notably, the MI \(\rightarrow\) CS transition becomes rather independent of the primary lattice depth, indicating that the crystalline state is a genuine many-body effect exclusively triggered by the long-range interactions.

Figs.~\ref{fig:PD-commensurate}(b)-(d) illustrate how correlated disorder modifies the physical landscape, as reflected in the changes in \(\Delta\), which reveal the destabilization of different quantum states in the periodic lattice.
The SF-MI phase transition observed in the periodic system (diagonal, yellow to purple region in Fig.~\ref{fig:PD-commensurate}) is strongly affected by correlated disorder.
Larger values of $V_d$ push the superfluid region to both deeper primary lattices (all the way to $V_p=16 E_r$ at weak interactions) and slightly stronger interaction strengths.
At the same time, this transition becomes progressively insensitive to the depth of the lattice, and only triggered by the increasing repulsions [see Fig.~\ref{fig:PD-commensurate}(b)-(d)]. 
On the other hand, the crystal state appearing at strong interactions -- $g_d \gtrsim 5 E_r$ -- is completely unaffected by the quasiperiodicity at the scale we probed, suggesting very strong robustness.

\begin{figure}
\centering
\includegraphics[width=1.0\columnwidth]{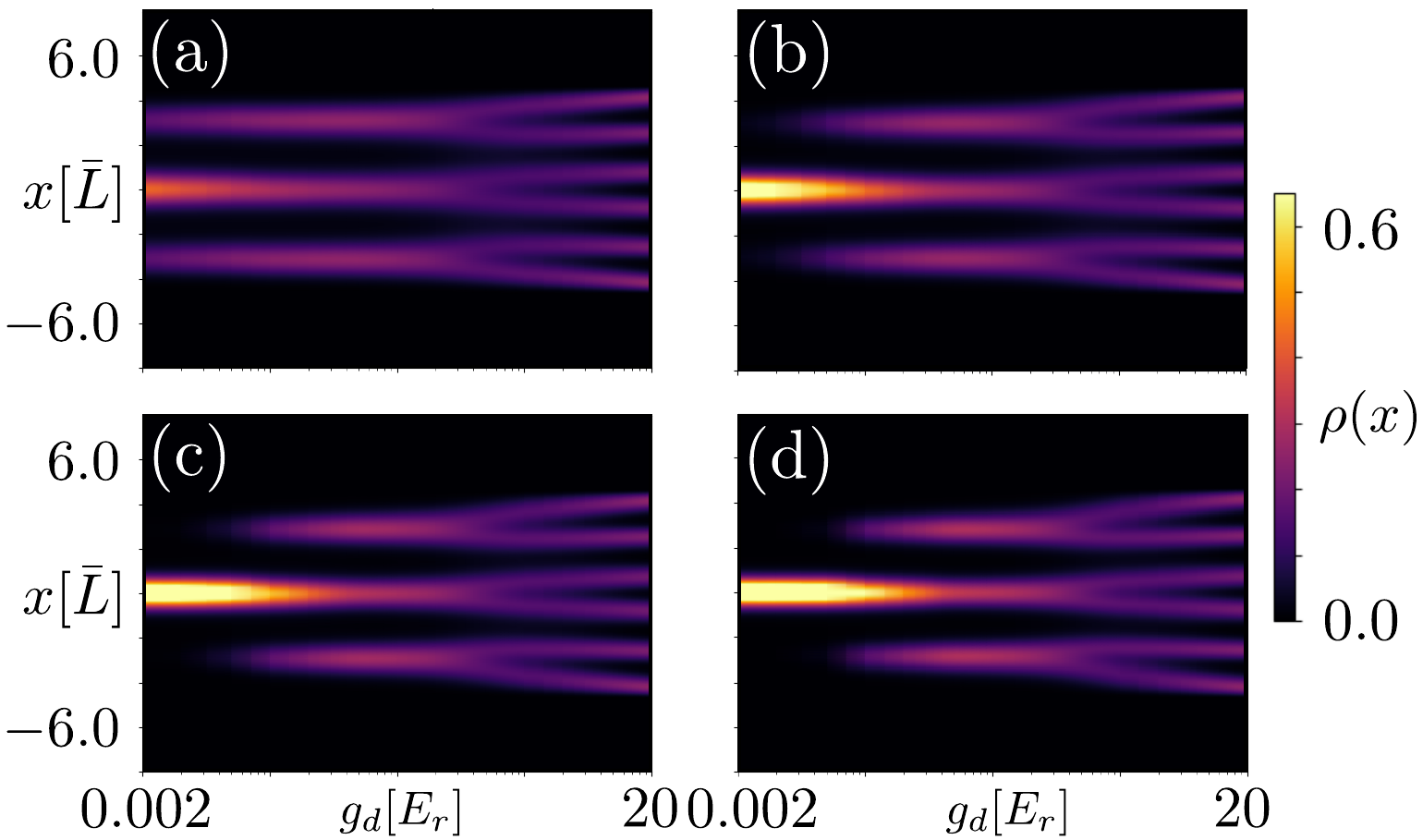}
\caption{Behavior of the real-space density $\rho(x)$ for the commensurate filling case $N=6$, $S=3$ with primary lattice depth $V_p=6 E_r$.
The four panels show results for four different disorder potential depths:
(a) $V_d=0.0 E_r$,
(b) $V_d=2.0 E_r$,
(c) $V_d=4.0 E_r$,
(d) $V_d=6.0 E_r$.
}
\label{fig:density-commensurate-Vp-6}
\end{figure}

\begin{figure}[h!]
\centering
\includegraphics[width=1.0\columnwidth]{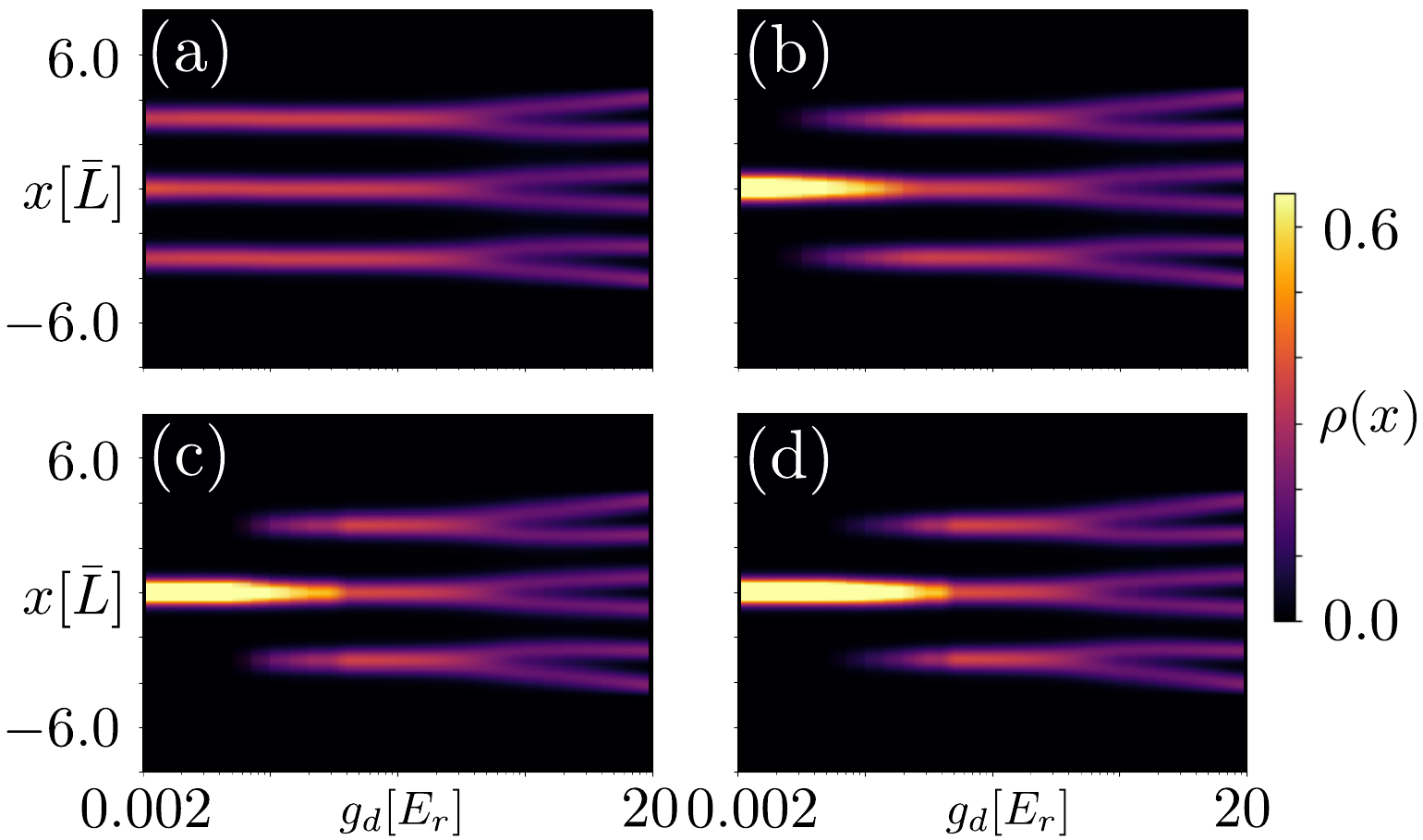}
\caption{Behavior of the real-space density $\rho(x)$ for the commensurate filling case $N=6$, $S=3$ with primary lattice depth $V_p=16 E_r$.
The four panels show results for four different disorder potential depths:
(a) $V_d=0.0 E_r$,
(b) $V_d=2.0 E_r$,
(c) $V_d=4.0 E_r$,
(d) $V_d=6.0 E_r$.
}
\label{fig:density-commensurate-Vp-16}
\end{figure}

\subsubsection{Density}
We provide further details on how the spatial differences among the various phases are impacted by correlated disorder by plotting the real-space density distribution $\rho(x)$ for a shallow lattice with $V_p=6.0 E_r$ [in Fig.~\ref{fig:density-commensurate-Vp-6}] and a deep lattice with $V_p=16.0 E_r$ [in Fig.~\ref{fig:density-commensurate-Vp-16}].
Each figure showcases how the density changes as a function of interaction strength $g_d$ under progressively stronger quasiperiodicity.

In the shallower lattice, the ground state at very weak interactions is a superfluid with a strong central peak and two small side peaks (at even smaller interactions, a more localized superfluid is recollected, see appendix~\ref{app:ultraweak}).
Given that this state exhibits global coherence while maintaining the crystalline structure imposed by the primary lattice, it is sometimes referred to as \emph{supersolid} in the literature~\cite{Pollet:2010,Boettcher:2019,Tanzi:2019,Chomaz:2019,Guo:2019,Natale:2019,Tanzi:2019-2,Tanzi:2021,Norcia:2021,Sohmen:2021,Sanchez-Baena:2023,Recati:2023}.
Due to the small size of our system, though, we refrain from explicitly using this terminology here.

In the deeper lattice, the ground state at very weak interactions is already a MI, as we can see from the three clearly delineated peaks.
Apart from the different initial state at weak interactions, the progression of the density profile as the interactions are increased is the same for both lattice depths: the peaks become progressively more localized (MI) until dimerization occurs, which leads the system into a CS with well-separated peaks.
Both transitions (SF $\to$ MI and MI $\to$ CS) are achieved gradually. 
We also remark that the CS explicitly breaks out of the geometric constraints imposed by the optical lattice, indicating that it cannot be described by standard lattice models like the Bose-Hubbard model.
The role of correlated disorder is also clear from the density.
While $V_d$ has a strong impact at weak interactions, reinforcing superfluidity [even nonmonotonically as shown in the deeper lattice -- Fig.~\ref{fig:density-commensurate-Vp-16}(b)], the CS occurring at strong interactions is unaffected.

\subsubsection{Energetics}
\begin{figure}[h!]
\centering
\includegraphics[width=1.0\columnwidth]{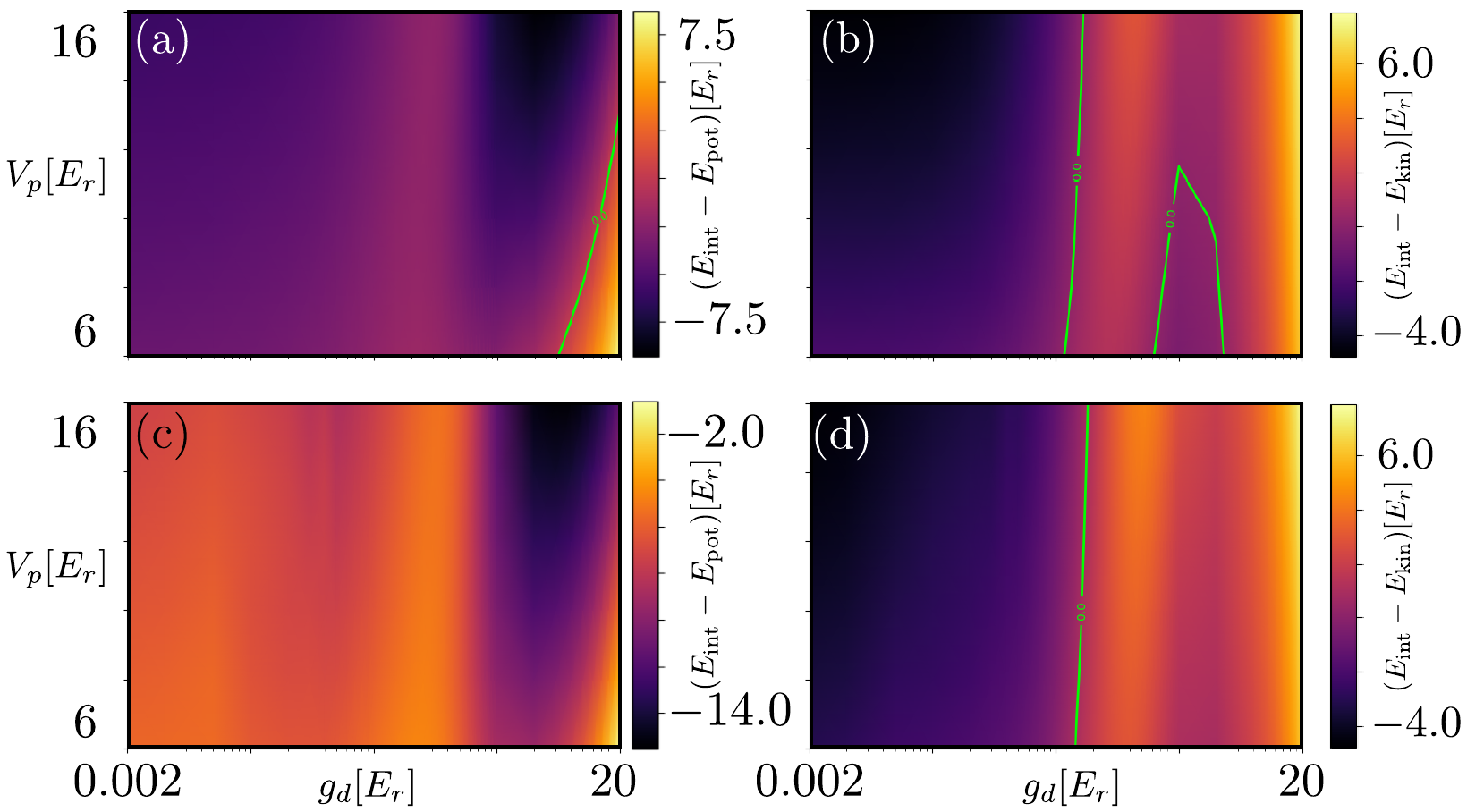}
\caption{Competition between different energy scales in the commensurate filling case $N=6$, $S=3$. 
(a) $E_{\mathrm{int}} - E_{\mathrm{pot}}$, $V_d=0.0 E_r$,
(b) $E_{\mathrm{int}} - E_{\mathrm{kin}}$, $V_d=0.0 E_r$,
(c) $E_{\mathrm{int}} - E_{\mathrm{pot}}$, $V_d=6.0 E_r$,
(d) $E_{\mathrm{int}} - E_{\mathrm{kin}}$, $V_d=6.0 E_r$,
}
\label{fig:energies-commensurate}
\end{figure}

Fig.~\ref{fig:energies-commensurate} provides further insights into the energetic stability of the different phases.
In it, we visualize the difference between interaction and potential energy (left panels) and interaction and kinetic energy (right panels) for the periodic case (top panels) and the disordered case (bottom panels).
The green lines further map out contour lines where this difference is zero.

As we can see, in most of the phase diagram the potential energy is the dominating energy scale, due to hard-wall boundary conditions.
Thus, the sequence of phase transitions is rather triggered by the interplay between kinetic and interaction energy.
In the SF phase, kinetic energy dominates.
As the system transitions towards MI, kinetic and interaction energies become comparable.
In the CS, instead, interaction energy is firmly the leading term. 
Moreover, the transition from MI to CS is characterized by a plateau in the total energy, suggesting that once the crystal state is formed, additional interaction energy does not significantly alter its structure.
This reinforces the observation that quasiperiodicity primarily affects intermediate phases while leaving the CS largely unaffected.
Interestingly enough, the periodic case exhibits a puddle around $g_d \approx 2 E_r$ where the kinetic energy becomes again slightly larger than the interaction energy.
However, we do not see any visible manifestations of this change in the density nor in the correlations.
This puddle disappears completely in the disordered case.

\subsubsection{Correlations and coherence}

\begin{figure}[h!]
\centering
\includegraphics[width=1.0\columnwidth]{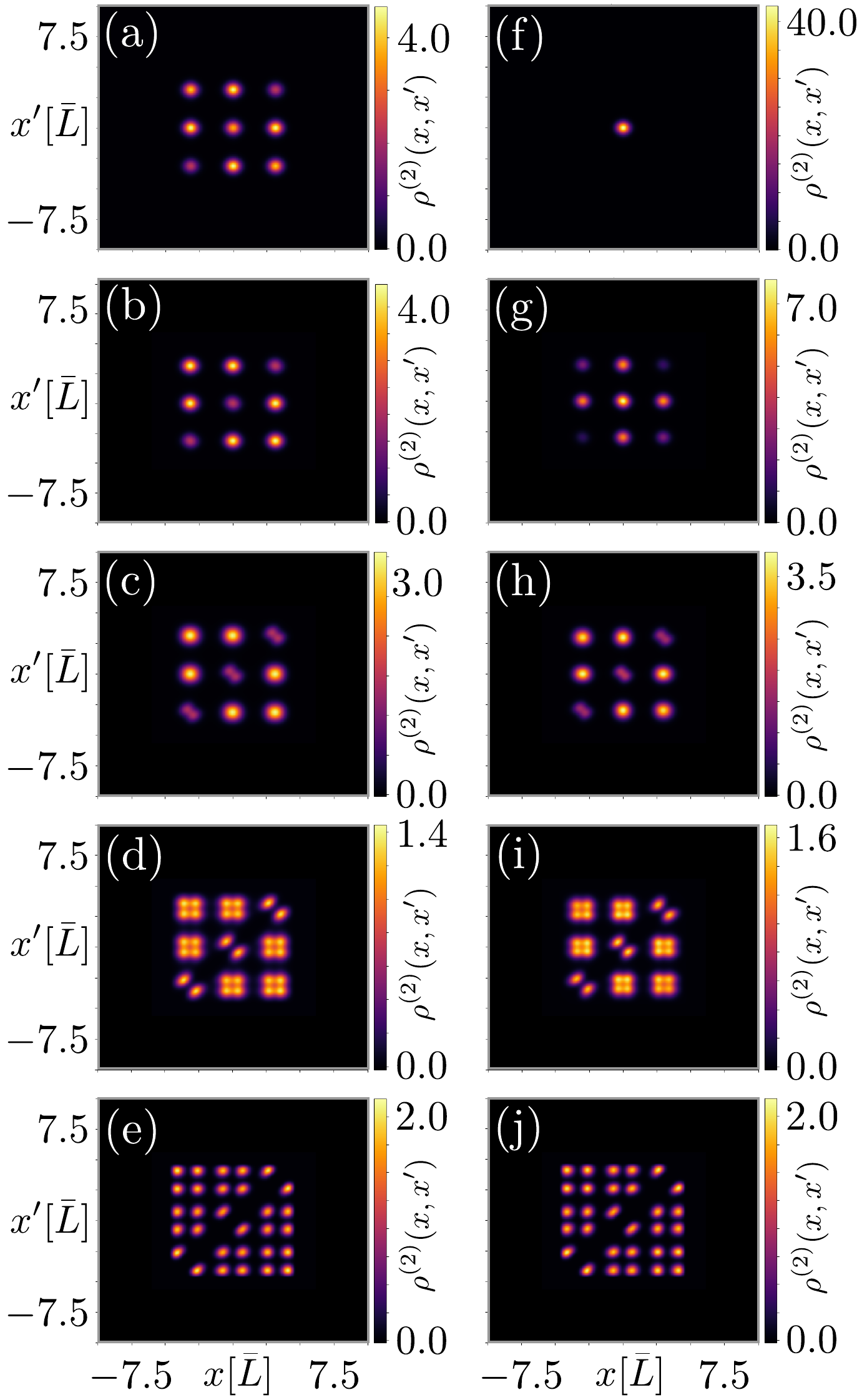}
\caption{Behavior of the 2-RDM $\rho^{(2)}(x,x')$ for commensurate filling. 
The panels show a comparison between the periodic case [$V_d=0.0 E_r$, panels (a)-(e)] and the quasiperiodic case  [$V_d=6.0 E_r$, panels (f)-(j)] for $V_p=16 E_r$ and across increasing interaction strengths $g_d$.
The values of the interactions are 
(a), (f): $g_d = 0.002 E_r$, 
(b), (g): $g_d = 0.08 E_r$, 
(c), (h): $g_d = 0.5 E_r$, 
(d), (i): $g_d = 1.6 E_r$, 
(e), (j): $g_d = 20.0 E_r$.
}
\label{fig:rho2-commensurate}
\end{figure}

To further clarify the density signature of the emergent phases in the disordered lattice, we compare the two-body density $\rho^{(2)}(x,x')$ between the periodic case and the quasiperiodic ($V_d=6.0 E_r$) case at $V_p=16.0 E_r$ for several values of the interaction strength $g_d$.
This is shown in Fig.~\ref{fig:rho2-commensurate}.
The effect of correlated disorder is striking in the Mott insulating phase at weak interactions [Fig.~\ref{fig:rho2-commensurate}(a) and (f)], where the entire state is collapsed to a single-site superfluid.
Even when interactions are increased [Figs.~\ref{fig:rho2-commensurate}(b) and (g)], there are still clearly more superfluid coherence patterns in the disordered case, where off-diagonal two-body coherence -- a hallmark of the MI state -- cannot be completely achieved.
As soon as the Mott dimerization occurs [Figs.~\ref{fig:rho2-commensurate}(c)-(e) and (h)-(j)], though, the correlation patterns become indistinguishable between periodic and quasiperiodic cases.

\subsection{Incommensurate filling}
\label{sec:results-incomm}

We now discuss our results pertaining to simulations of a system with incommensurate filling with $N=8$ particles and $S=5$ sites.
In this setting, SF, MI, and CS still exist.
However, the additional ingredient introduced by incommensuration leads to a more diverse landscape of quantum phases. 
Most notably, incommensuration effects can stabilize states with inhomogeneous population, like charge density waves (CDW), characterized by alternate filling of lattice sites~\cite{Hughes:2023, Bilinskaya:2024}.

\subsubsection{Order parameter}

\begin{figure}
\centering
\includegraphics[width=1.0\columnwidth]{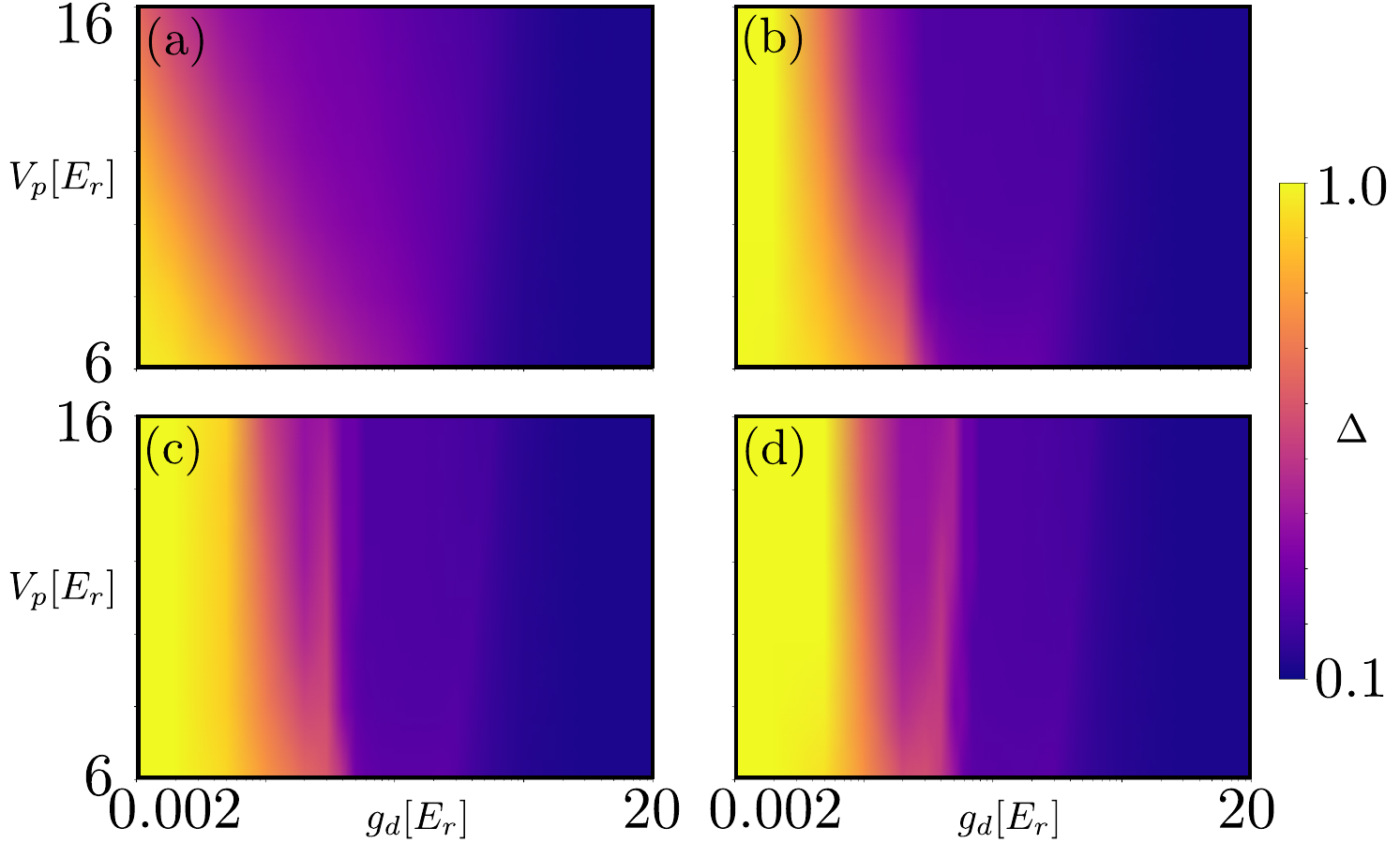}
\caption{Order parameter diagrams for the incommensurate filling case $N=8$, $S=5$. 
The plots showcase the order parameter $\Delta$ as a function of the primary lattice depth $V_p$ and the interaction strength $g_d$.
The four panels show results for four different disorder potential depths:
(a) $V_d=0.0 E_r$,
(b) $V_d=2.0 E_r$,
(c) $V_d=4.0 E_r$,
(d) $V_d=6.0 E_r$.
}
\label{fig:PD-incommensurate}
\end{figure}

In Fig.~\ref{fig:PD-incommensurate}, we visualize the behavior of the order parameter $\Delta$ in the $(g_d,V_p)$ parameter space and for increasing values of disorder strength $V_d$.
Overall, the trend is similar to the commensurate case.
In the periodic case [Fig.~\ref{fig:PD-incommensurate}(a)], for weak interactions and shallower lattices, the particles exist in a SF state with $\Delta \sim 1$ (yellow region).
Increasing the interactions or deepening the lattice pushes the system into more fragmented states with multiple-orbital population and $\Delta < 1$ (orange, red, and purple regions).
As we shall see later, these states can be MI or CDW depending on the lattice depth, but cannot be easily recognized from $\Delta$ alone since they can both be described by the same number of orbitals leading to $\Delta = \frac{1}{S}=\frac{1}{5}$.
From around $g_d \sim 0.2 E_r$ onward, the dependence on the potential depth becomes drastically weaker and the decrease in $\Delta$ is mostly dictated by the increase in interaction strength, which finally pushes the particles into the crystal state with $\Delta=\frac{1}{N}=\frac{1}{8}$ (dark blue region).

When the quasiperiodicity is turned on [Fig.~\ref{fig:PD-incommensurate}(b)-(d)], much of the weak-interaction regime becomes dominated by the SF phase, which also engulfs regions that were previously fragmented up to $g_d \sim 0.05 E_r$.
The SF also conquers previously MI regions at large values of $V_p$.
The CS remain instead rather unaffected by disorder.

\subsubsection{Imbalance}

In Ref.~\cite{Chatterjee:2020}, it has been claimed that the periodic system actually exhibits two kinds of CS ordering, depending on the strength of the interparticle interactions.
A first CS -- dubbed kinetic crystal state (KCS) due to the kinetic energy being the dominant term -- forms at intermediate repulsions and is characterized by a full split of only the (three) innermost peaks resulting in an occupation pattern $[1,2,2,2,1]$ for $N=8$ particles.
At a critical threshold of $g_c$, a structural reconfiguration suddenly takes place, whereupon the bosons arrange themselves in a CDW pattern with occupations $[2,1,2,1,2]$.
This type of crystal state is then termed ``density-wave crystal state" (DWCS).

\begin{figure}
\centering
\includegraphics[width=1.0\columnwidth]{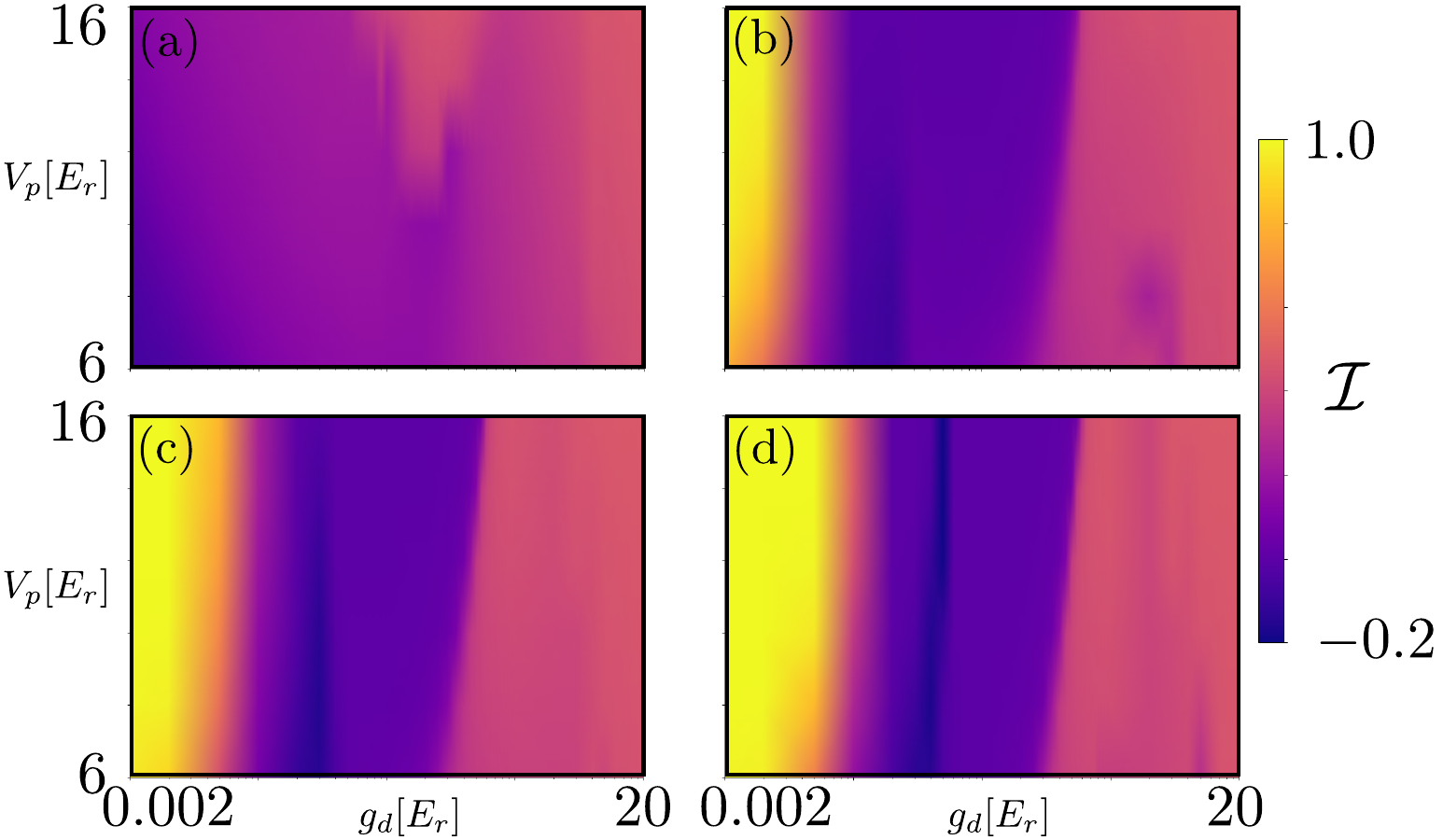}
\caption{Behavior of the imbalance $\mathcal{I}$ for the incommensurate filling case $N=8$, $S=5$. 
The observable is calculated a function of the primary lattice depth $V_p$ and the interaction strength $g_d$.
The four panels show results for four different disorder potential depths:
(a) $V_d=0.0 E_r$,
(b) $V_d=2.0 E_r$,
(c) $V_d=4.0 E_r$,
(d) $V_d=6.0 E_r$.
}
\label{fig:imbalance-incommensurate}
\end{figure}

To verify these claims and inspect the fate of these putative CS flavors under correlated disorder, we present a systematic scan of the imbalance $\mathcal{I}$ over the entire parameter space in Fig.~\ref{fig:imbalance-incommensurate}.
The imbalance should by its nature detect any discontinuity in the population distribution across different sites.
Surprisingly, our findings show that -- while we do observe quantitative changes in the imbalance for different phases -- there is no clear signal that would distinguish different types of CS over extended regions in parameter space in the periodic case.
Instead, the imbalance progressively increases as interactions become strongers and the lattice deeper.
This suggests that there is no evidence of KCS / DWCS difference in the periodic system, and the findings of Ref.~\cite{Chatterjee:2020} might have been a numerical artifact caused by a insufficient number of orbitals or too narrow simulation grids~\footnote{In MCTDH-X, the system is defined for periodic boundary conditions (PBC) by default. Hard-wall boundaries have to be imposed separately on top of the PBC. If the physical boundaries (hard walls) are too close to the simulation boundaries, self-interaction effects across the PBC might occur.}
There is a notable exception to this behavior, indicated by the red bubble around $g_d \sim 0.4 E_r$ and $V_p > 9 E_r$, which however indicates a different pathway for the MI $\to$ CS transition and not a KCS.
We shall return to as we discuss the density distribution below.

In the quasiperiodic case, Figs.~\ref{fig:imbalance-incommensurate}(b)-(d) highlight a strong qualitative change to the imbalance diagram.
The collapse into single-site localized SF at weak interactions maximizes the imbalance to 1 (yellow region).
Similarly to the order parameter $\Delta$, this region stretches to deep optical lattices with $V_p=16 E_r$.
As interactions are increased, the imbalance undergoes three changes, which are only weakly dependent on the lattice depth.
The first is a smooth crossover to lower values as the state transitions from SF to MI.
The second is a temporary dip in the imbalance in concomitance with the spread of the MI state to outer sites (see next section).
The third is a rather sudden increase in imbalance corresponding to the transition from MI to CS that manifests as a dimerization process in the density.
Increasing the disorder pushes the onset of the first two transitions to higher interactions.
However, the third transition to CS is unperturbed, reinforcing the stability of this phase.

\subsubsection{Density}
\begin{figure}
\centering
\includegraphics[width=1.0\columnwidth]{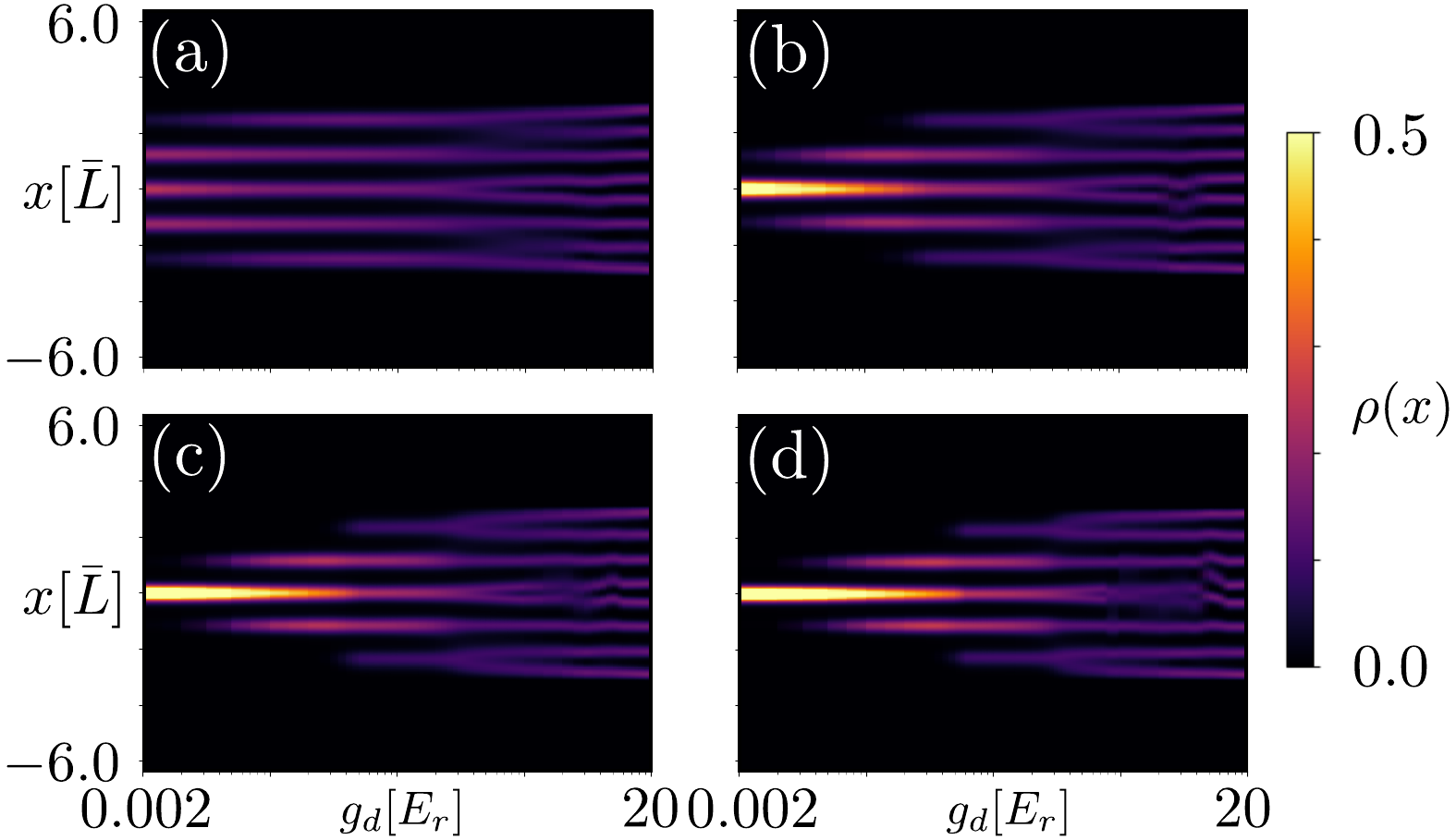}
\caption{Real-space density $\rho(x)$ as a function of interaction strength $g_d$ for the incommensurate filling case $N=8$, $S=5$ with primary lattice depth $V_p=6 E_r$.
The four panels show results for four different disorder potential depths:
(a) $V_d=0.0 E_r$,
(b) $V_d=2.0 E_r$,
(c) $V_d=4.0 E_r$,
(d) $V_d=6.0 E_r$.
}
\label{fig:density-incommensurate-Vp-6}
\end{figure}

\begin{figure}
\centering
\includegraphics[width=1.0\columnwidth]{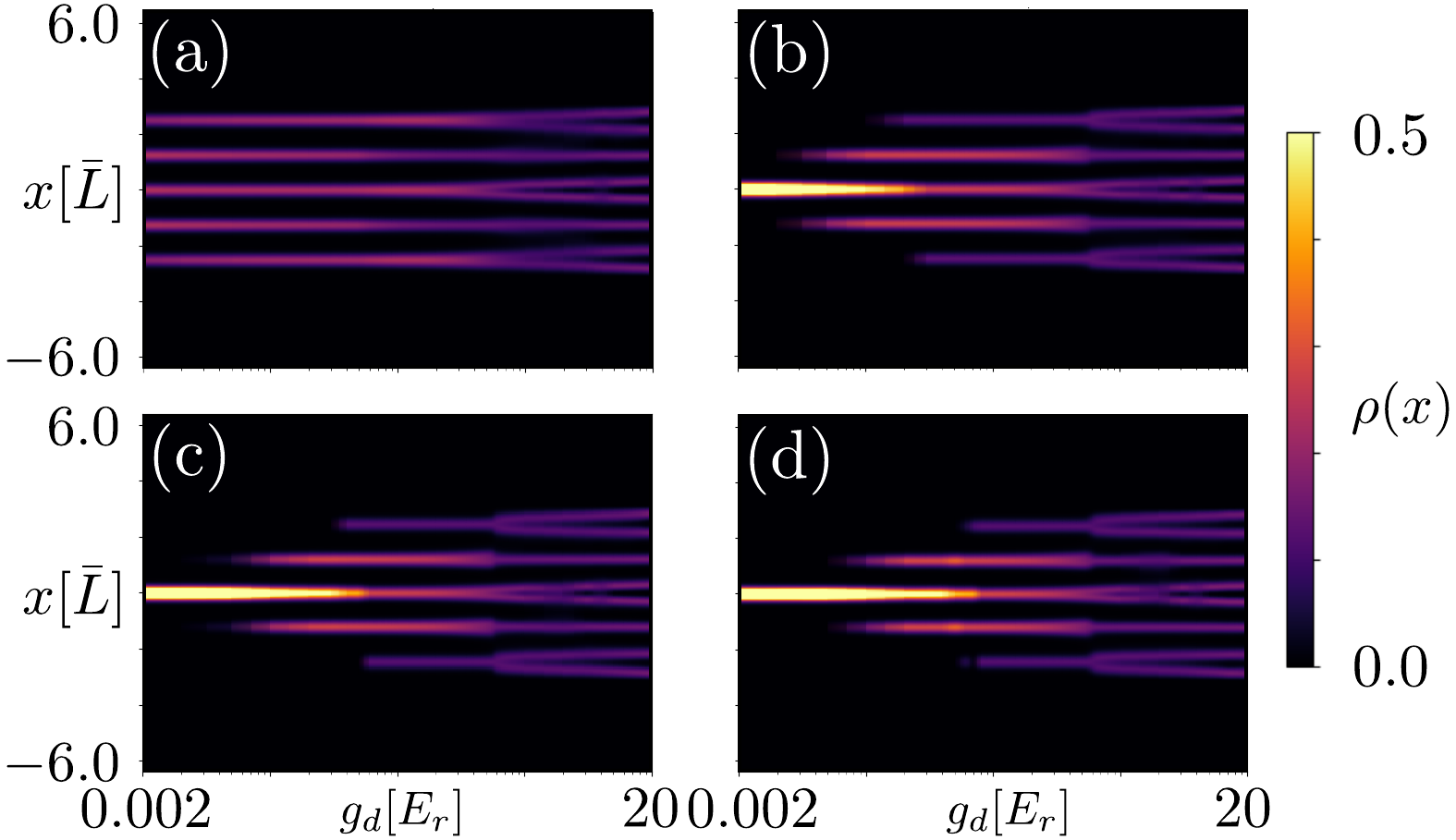}
\caption{Real-space density $\rho(x)$ as a function of interaction strength $g_d$ for the incommensurate filling case $N=8$, $S=5$ with primary lattice depth $V_p=16 E_r$.
The four panels show results for four different disorder potential depths:
(a) $V_d=0.0 E_r$,
(b) $V_d=2.0 E_r$,
(c) $V_d=4.0 E_r$,
(d) $V_d=6.0 E_r$.
}
\label{fig:density-incommensurate-Vp-16}
\end{figure}

To better address the nature of the different phases that the imbalance can track, we visualize the real-space density distribution as a function of $g_d$ and increasing disorder strength $V_d$ for a shallower ($V_p=6 E_r$, Fig.~\ref{fig:density-incommensurate-Vp-6}) and deeper ($V_p=16 E_r$, Fig.~\ref{fig:density-incommensurate-Vp-16}) lattice, respectively.

\begin{figure}
\centering
\includegraphics[width=1.0\columnwidth]{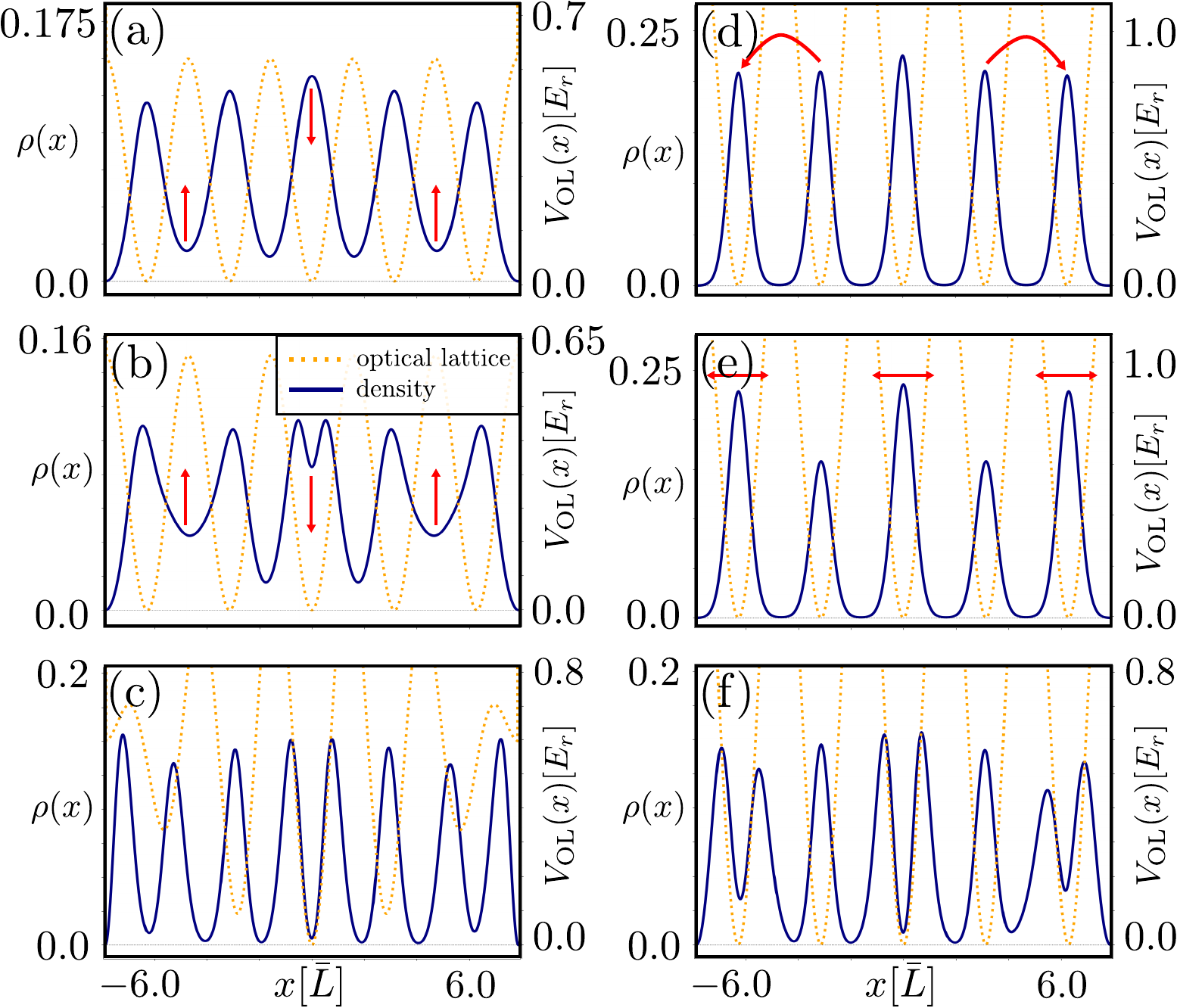}
\caption{Emergence of nonhomogeneous profiles in the real-space density $\rho(x)$ for the incommensurate filling case $N=8$, $S=5$ and periodic system $V_d=0.0 E_r$.
The panels show the different behavior in the shallower lattice with
(a) $V_p=6.0 E_r$, $g_d=0.4 E_r$,
(b) $V_p=6.0 E_r$, $g_d=1.0 E_r$, and
(c) $V_p=6.0 E_r$, $g_d=14.0 E_r$,
and the deeper lattice with
(d) $V_p=16.0 E_r$, $g_d=0.02 E_r$ 
(e) $V_p=16.0 E_r$, $g_d=0.2 E_r$, and
(f) $V_p=16.0 E_r$, $g_d=10.0 E_r$.
The red arrows indicate the population transfer across the transitions.
}
\label{fig:density-incommensurate-clean}
\end{figure}

In the shallower lattice, the initial state at weak interactions [cf. Figs.~\ref{fig:density-incommensurate-Vp-6}(a)] is a SF delocalized over three lattice sites (or ``supersolid", see discussion in Sec.~\ref{sec:results-comm}), since the order parameter in this region is $\Delta \sim 1$.
In the periodic potential, stronger interactions push this initial configuration into a MI which persists over many orders of magnitude.
A CS eventually sets in as interactions are increased further.
However, the mechanism is strongly dependent on the primary lattice depth $V_p$, as further illustrated in Fig.~\ref{fig:density-incommensurate-clean}.
In a shallower lattice [Fig.~\ref{fig:density-incommensurate-clean}(a)-(b)], the MI first transitions into an intermediate state where the central site hosts two particles that dimerize, while the two outermost sites share three particles each, with an increase in intersite density around $x \sim \pm 5.5 \bar{L}$.
This is the approximate location where the a third peak eventually emerges in the full CS [Fig.~\ref{fig:density-incommensurate-clean}(c)].

In the deeper lattice, the initial state at weak interactions [cf. Figs.~\ref{fig:density-incommensurate-Vp-16}(a)] is instead already a full-fledged MI with approximately equal population on all the five lattice sites.
As interactions are increased, first the Mott localization simply increases.
Then, however, the density reorganization pushes the system into an intermediate CDW state with a population transfer from the second-outermost to the outermost peaks [Fig.~\ref{fig:density-incommensurate-clean}(d)-(e)].
This leads to a [2,1,2,1,2] pattern in the density. 
Upon entering the CS state, only the odd lattice sites dimerize and bring about the crystalline pattern the breaks away from the optical lattice structure [Fig.~\ref{fig:density-incommensurate-clean}(f)].
This pattern is similar to the DWCS observed in Ref.~\cite{Chatterjee:2020}, albeit without a preliminary KCS and with interpeak spacing converging to a uniform value as interactions reach $g_d \to 20 E_r$~\footnote{We remark that the slight wiggles observed in both the density heatmaps and the density profiles are not true features but likely small numerical imprecisions due to convergence difficulties when many strongly interacting particles are packed in a restricted space as in our simulations.}.

\begin{figure}
\centering
\includegraphics[width=1.0\columnwidth]{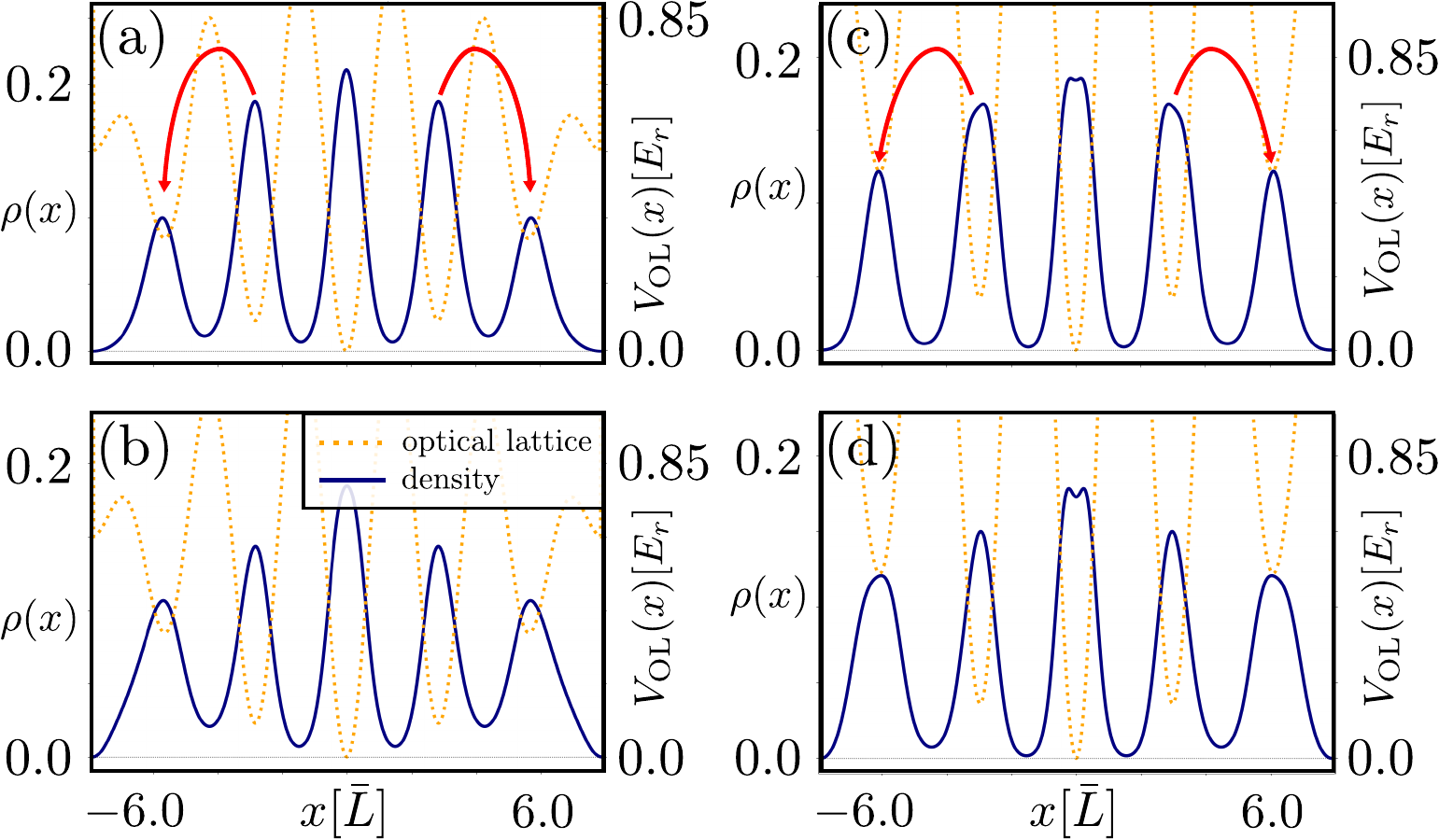}
\caption{Emergence of CDW ordering in the real-space density $\rho(x)$ for the incommensurate filling case $N=8$, $S=5$ and strong disorder $V_d=6.0 E_r$.
The panels show different values of primary lattice depth and interaction strength:
(a) $V_p=6.0 E_r$, $g_d=0.4 E_r$,
(b) $V_p=6.0 E_r$, $g_d=0.6 E_r$,
(c) $V_p=16.0 E_r$, $g_d=1.1 E_r$,
(d) $V_p=16.0 E_r$, $g_d=1.2 E_r$.
The red arrows indicate the population transfer across the transition.
}
\label{fig:density-incommensurate-KCS}
\end{figure}

When correlated disorder is added to the incommensurate problem, the state at weak interactions is converted to a single-site SF as already indicated by the order parameter and imbalance plots.
This can be clearly observed for both shallower and deeper optical lattices in Figs.~\ref{fig:density-incommensurate-Vp-6}(b)-(d) and ~\ref{fig:density-incommensurate-Vp-16}
(b)-(d).
Upon increasing the interaction strength $g_d$, the usual transitions to MI and later to CS take place.
However, the smooth transition from MI to CS without the intermediate CDW phase -- as observed in the shallower periodic lattices -- is suppressed.
Instead, the system always reaches the CS by passing through a CDW regime where density is redistributed towards the outermost peaks from their immediate neighbors, as shown in Fig.~\ref{fig:density-incommensurate-KCS}.
Interestingly, the disordered system does exhibit a [1,2,2,2,1] density configuration in the MI phase before transitioning to the CDW phase [cf. Fig.~\ref{fig:density-incommensurate-KCS}(a),(c)]. 
However, as opposed to the results of Ref.~\cite{Chatterjee:2020}, the innermost peaks are never dimerized.
Instead, the CDW pattern is achieved via population transfer between separate peaks [cf. Fig.~\ref{fig:density-incommensurate-KCS}(b),(d)].
This indicates the absence of the KCS phase even in the quasiperiodic case.

\subsubsection{Energetics}

\begin{figure}[h!]
\centering
\includegraphics[width=1.0\columnwidth]{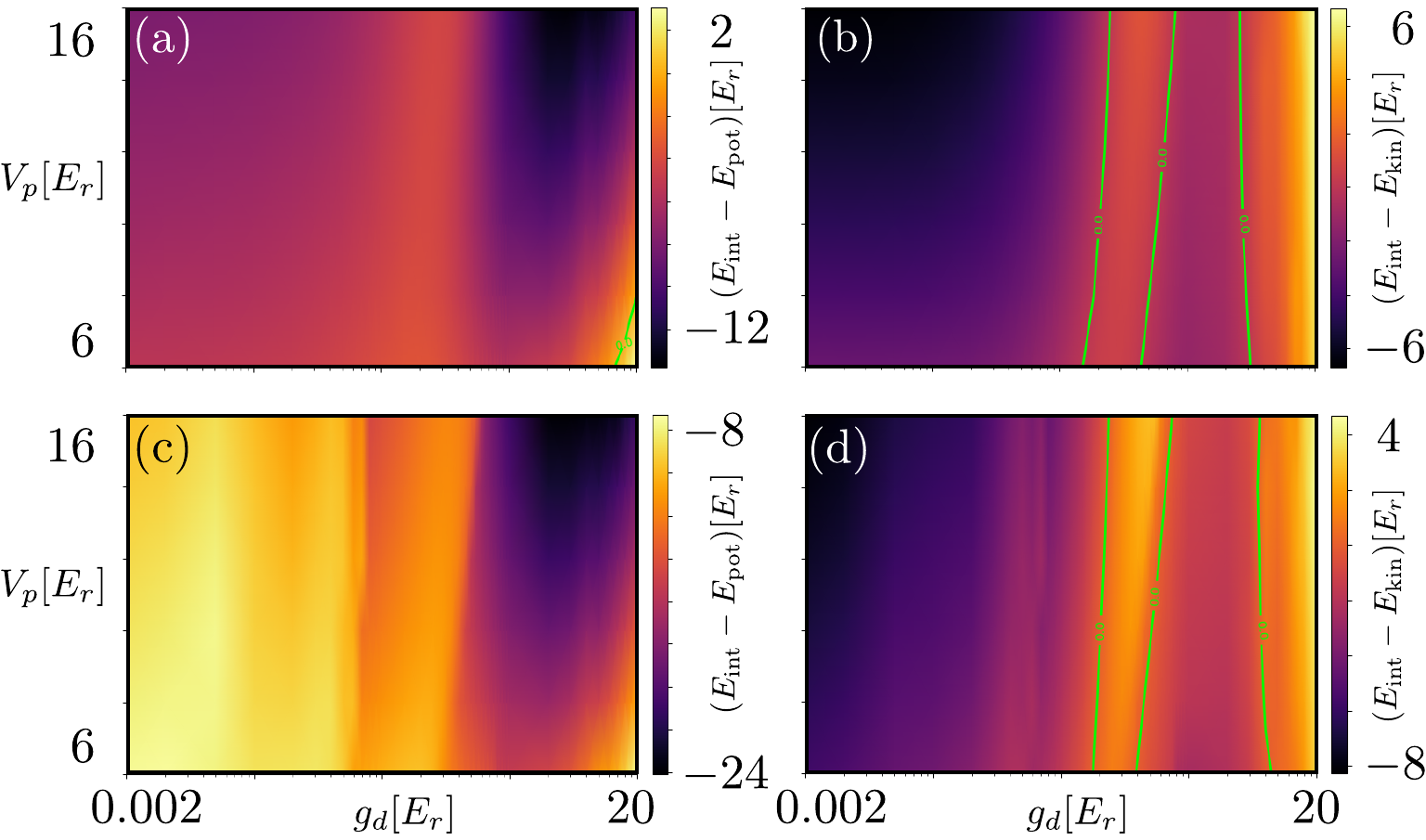}
\caption{Competition between different energy scales in the incommensurate filling case $N=8$, $S=5$. 
(a) $E_{\mathrm{int}} - E_{\mathrm{pot}}$, $V_d=0.0 E_r$,
(b) $E_{\mathrm{int}} - E_{\mathrm{kin}}$, $V_d=0.0 E_r$,
(c) $E_{\mathrm{int}} - E_{\mathrm{pot}}$, $V_d=6.0 E_r$,
(d) $E_{\mathrm{int}} - E_{\mathrm{kin}}$, $V_d=6.0 E_r$.
The green lines represent zero-contour lines where the two energy scales are equal.
}
\label{fig:energies-incommensurate}
\end{figure}

We now discuss how the different phases appearing in the system with incommensurate filling differ in terms of energetics.
Like in the commensurate case, we find that the potential energy is the dominant energy scale in the vast majority of the parameter space.
Therefore, Fig.~\ref{fig:energies-incommensurate} visualizes only the difference between interaction and potential energy [panels (a),(c)] and between interaction and kinetic energy [panels (b),(d)] for the periodic case [panels (a),(b)] and the strongly disordered case with $V_d=6.0 E_r$ [panels (c),(d)].

Qualitatively, the behavior of the different energy components is similar to that of the commensurate case at weak interactions, but exhibits notable differences at high interactions due to the presence of a CDW state.
In the SF region, the kinetic energy clearly dominates over the interaction case.
This dominance is progressively reduced with increasing interaction strength, and upon entering the MI region the interaction energy becomes more dominant (around $g_d \sim 0.5 E_r$). 
However, when the transition to CDW state takes place, the energy trend reverses and the kinetic energy becomes again larger than the interaction energy (around $g_d \sim 1.0 E_r)$.
This change reinforces the mechanism of particle redistribution via hopping to the outermost sites already outlined in the discussion of the density results.
In retrospect, then, we can characterize the similar pool of increased kinetic energy observed in the commensurate case [cf. Fig.~\ref{fig:energies-commensurate}(b)] as a potential precursor of a many-body phase where particles are more mobile, as in the CDW phase. 
Once the interaction strength reaches the level needed to trigger a transition to the CS phase (around $g_d \sim 10 E_r$), the trends reverse one more time and the interaction energy becomes again the dominant term.

The introduction of quasiperiodicity does not alter the overall energetics pattern observed in the periodic case.
Remarkably, and in contrast to the commensurate case, the quasiperiodicity does not affect the nonmonotonic behavior of kinetic and interaction energies.
This can be justified with the observation that the CDW state is not destroyed by correlated disorder.

\subsubsection{Correlation and coherence}

\begin{figure}
\centering
\includegraphics[width=1.0\columnwidth]{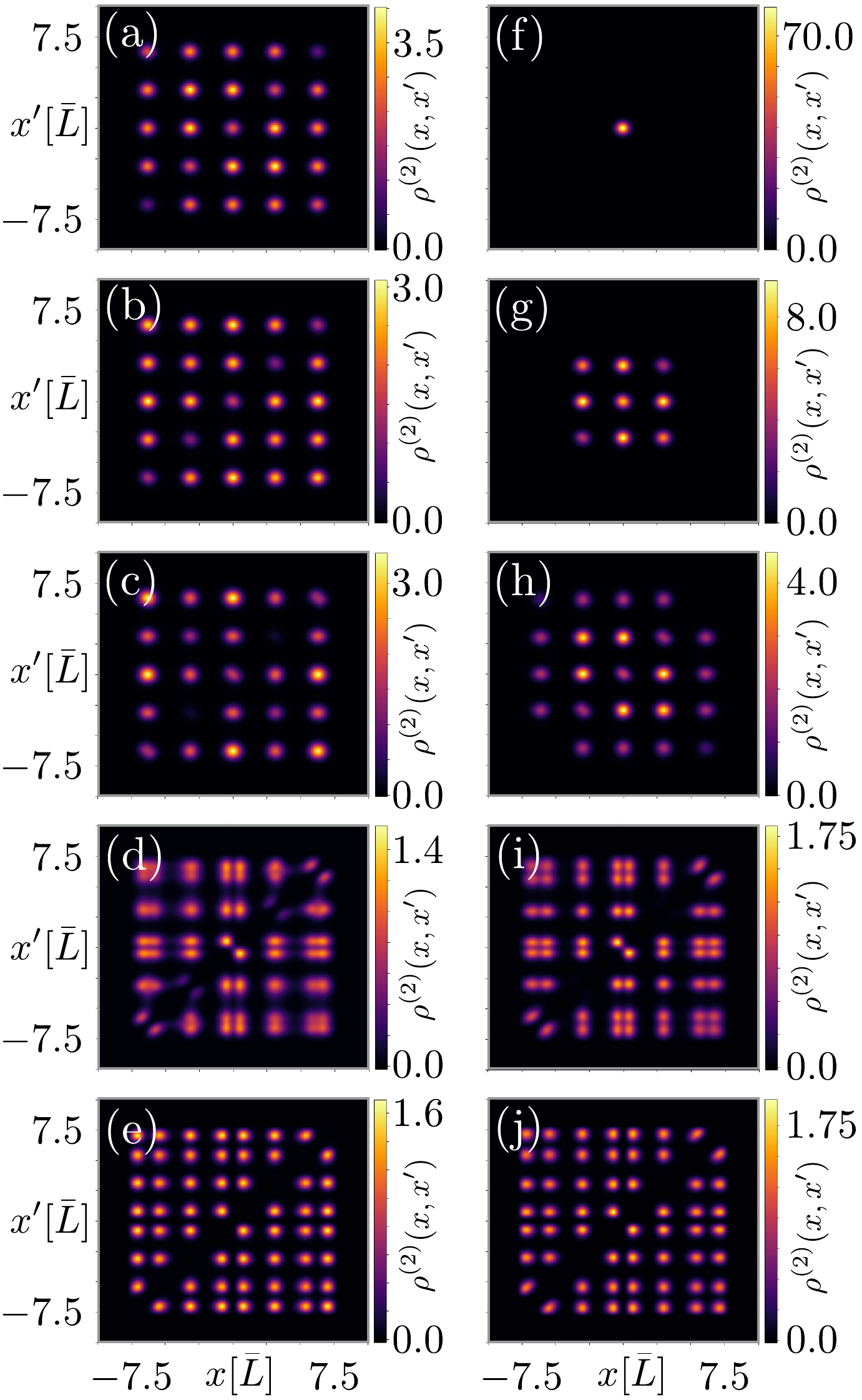}
\caption{Behavior of the 2-RDM $\rho^{(2)}(x,x')$ for incommensurate filling. 
The panels show a comparison between the periodic case [$V_d=0.0 E_r$, panels (a)-(e)] and the quasiperiodic case  [$V_d=6.0 E_r$, panels (f)-(j)] for $V_p=16 E_r$ and across increasing interaction strengths $g_d$.
The values of the interactions are 
(a), (f): $g_d = 0.002 E_r$, 
(b), (g): $g_d = 0.1 E_r$, 
(c), (h): $g_d = 0.3 E_r$, 
(d), (i): $g_d = 2.0 E_r$, 
(e), (j): $g_d = 20.0 E_r$.
}
\label{fig:rho2-incommensurate}
\end{figure}

We conclude our analysis of the incommensurate filling case by examining the behavior of the 2-RDM for a selection of five different interaction strengths across the periodic and quasiperiodic case for the deeper primary lattice computations with $V_p=16 E_r$.
The results are shown in Fig.~\ref{fig:rho2-incommensurate}.
In the periodic case, the system at weak interactions exhibits both on- and off-diagonal second-order correlation [cf. Fig.~\ref{fig:rho2-incommensurate}(a)] -- in particular within the central three sites -- which are the hallmarks of SF and MI, respectively.
When the repulsions are increased, the $\rho^{(2)}$ profile follows more strongly that of a MI, with weakening diagonal correlation [cf. Fig.~\ref{fig:rho2-incommensurate}(b)].
Then, in Fig.~\ref{fig:rho2-incommensurate}(c), the system loses diagonal correlation in the even lattice sites while maintaining strong on- and off-diagonal correlation across the odd sites, indicating the formation of the CDW.
Further increasing the interactions leads to a split of the doubly-occupied sites [cf. Fig.~\ref{fig:rho2-incommensurate}(d)], which eventually converge to a correlation pattern away from the optical lattice minima and without diagonal coherence [cf. Fig.~\ref{fig:rho2-incommensurate}(e)], indicating the presence of the CS.

In the presence of correlated disorder, the states at weak interactions display a completely different correlation pattern due to their SF nature, with a single populated central site [cf. Fig.~\ref{fig:rho2-incommensurate}(f)].
The effect of disorder persists until $g_d \sim 0.3 E_r$, and effectively delays the onset of the MI character [cf. Fig.~\ref{fig:rho2-incommensurate}(g) and (h)] and CDW to stronger interactions.
Eventually, however, the dimerized CDW [cf. Fig.~\ref{fig:rho2-incommensurate}(i)] and final CS [cf. Fig.~\ref{fig:rho2-incommensurate}(j)] display the exact same correlation pattern observed in the periodic case, further reinforcing the robustness of this interaction-induced many-body state of matter.

\section{Conclusions} 
\label{sec:conclusions}

In the commensurate filling scenario, our study has uncovered a rich interplay between dipolar interactions, kinetic energy, and quasiperiodic disorder. 
At weak interactions, we observe a transition from a superfluid (SF) state to a Mott insulator (MI) as the lattice depth increases. 
This transition is governed by the competition between kinetic energy and the external potential. As interactions strengthen, a density-modulated crystal state (CS) emerges, characterized by the spatial separation of particles into distinct peaks that are displaced from the lattice minima due to the long-range repulsive forces. 
Importantly, our results demonstrate that, while the SF and MI are highly sensitive to quasiperiodic disorder, the CS phase is instead very robust, maintaining its structural integrity even under strong disorder strengths. 
This highlights the fundamental role of dipolar interactions in stabilizing highly ordered quantum phases beyond traditional short-range interaction models.

For incommensurate fillings, our study reveals the emergence of additional phases beyond those observed in the commensurate case. 
Specifically, in addition to SF, MI, and CS phases, we identify the formation of charge density wave (CDW) states at intermediate interaction strengths. 
These states arise due to the combined effects of incommensurate lattice geometry and dipolar interactions, leading to an uneven occupation pattern across the lattice sites. 
Interestingly, the transition from MI to CS in the incommensurate system proceeds through this intermediate CDW phase irrespective of the disorder strength, while it is absent in the commensurate case even in the periodic case. 
Moreover, our analysis disproves the presence of a kinetic crystal state (KCS) suggested in previous studies.
While we observe a nonmonotonic resurgence of kinetic energy dominance in the CDW phase, we do not find any density splitting into single-particle peaks, which is a defining feature of the CS.

Our findings provide significant insights into the stability and dynamics of dipolar bosons in quasiperiodic optical lattices, with implications for both theoretical and experimental research.
Theoretically, our results enrich the understanding of long-range interacting quantum systems by identifying the resilience of dipolar crystal states under quasiperiodic disorder. 
This robustness suggests that dipolar crystals could serve as an ideal platform for studying many-body localization effects in correlated systems~\cite{Schreiber:2015,Bordia:2017}. 
Combining quasiperiodicity with long-range interactions is expected to give rise to more exotic localization mechanisms, such as Bose glass phases~\cite{Yu:2024} or dynamically self-bound states~\cite{Molignini:2025-quasicryst}.
Furthermore, the identification of the CDW state in incommensurate lattices opens new avenues for exploring unconventional insulating phases, which may have analogs in strongly correlated electron systems and quantum Hall physics~\cite{Roux:2008, Yu:2021}.

From an experimental perspective, our predictions are well within reach of current ultracold atomic experiments. 
State-of-the-art techniques allow for the realization of quasiperiodic optical lattices with precisely tunable disorder strengths, and recent advancements in controlling dipolar gases -- such as erbium and dysprosium atoms~\cite{Aikawa:2012, Lu:2011} -- provide an ideal testbed for our theoretical predictions. 
Our study suggests clear experimental signatures, such as density redistribution patterns and correlations, which can be directly measured using quantum gas microscopy~\cite{Gross:2021}. 
This paves the way for future investigations into dipolar many-body physics and the realization of novel quantum phases in engineered optical potentials.
Moreover, the impressive stability of crystal states could make them very suitable in quantum information processing platforms~\cite{DeMille:2002, Rabl:2007, Weimer:2012, Islam:2013}.

Our work opens several exciting directions for future research. 
One immediate extension would be the exploration of dynamical effects, such as the response of dipolar bosons to sudden changes in interaction strength or disorder potential~\cite{Molignini:2024,Molignini:2024-2}, which could shed light on nonequilibrium phase transitions and many-body localization phenomena. 
Another promising avenue is the study of various kinds of perturbations, from periodic drives to finite-temperature effects, which could provide crucial insights into the structural stability of the identified quantum phases and the role of thermal fluctuations in melting dipolar crystals, similar to the methods employed to study the mechanism that leads to Pauli crystal melting~\cite{Xiang:2023}.
Additionally, since quasiperiodic patterns can be readily engineered in 2D and 3D configurations~\cite{Viebahn:2019, Sbroscia:2020}, future investigations could explore such higher-dimensional systems, where the interplay between quasiperiodicity and long-range interactions could give rise to even richer phase diagrams, including novel supersolid and stripe phases. 
Finally, extending our study to fermionic dipolar gases would offer a complementary perspective on the role of quantum statistics in shaping the competition between disorder and interactions. 
These studies would further our understanding of long-range interacting quantum systems and potentially uncover new paradigms for quantum simulation and quantum technology applications.


\textit{Acknowledgements --}
We thank Budhatiya Chatterjee and R. Chitra for useful discussions.
This work was supported by the Swedish Research Council under grant 2024-05213.
Computation time at the High-Performance Computing Center Stuttgart (HLRS), on the Euler cluster at the High-Performance Computing Center of ETH Zurich, and at the High-Performance Computing cluster of Stockholm University (SUNRISE) is gratefully acknowledged. 

\appendix
\section{MCTDH-X}
\label{app:MCTDHX}
In this appendix, we provide a concise overview of the numerical method used to compute the time evolution of the few-boson systems discussed in the main text. 
We utilize the MultiConfigurational Time-Dependent Hartree method for indistinguishable particles, implemented in the MCTDH-X software~\cite{Alon:2008, Lode:2016, Fasshauer:2016, Lin:2020, Lode:2020, Molignini:2025-SciPost, MCTDHX}. 
This method, designed to solve the many-body Schr\"{o}dinger equation over time, is particularly suited for investigating the ground state properties of interacting and disordered ultracold systems. 
Notably, MCTDH-X has been successfully applied to a range of dipolar and long-range interacting systems~\cite{Fischer:2015, Lode:2017, Lode.njp:2018, Molignini:2018, Chatterjee:2018, Chatterjee:2019, Bera:2019, Lin:2019, Lin2:2019, Chatterjee:2020, Lin:2020-PRA, Lin:2021,  Molignini:2022, Bilinskaya:2024, Molignini:2024, Molignini:2024-2, Molignini:2025-quasicryst}.

The MCTDH-X framework represents the many-body wave function as a time-dependent linear combination of permanents:
\begin{equation}
\left| \Psi(t) \right>= \sum_{\mathbf{n}}^{} C_{\mathbf{n}}(t)\vert \mathbf{n};t\rangle.
\label{many_body_wf}
\end{equation}
These permanents are constructed from $M$ single-particle orbitals, which are time-dependent functions, as follows:
\begin{equation}
\vert \mathbf{n};t\rangle = \prod^M_{k=1}\left[ \frac{(\hat{b}_k^\dagger(t))^{n_k}}{\sqrt{n_k!}}\right] |0\rangle,
\label{many_body_wf_2}
\end{equation}
where $\mathbf{n}=(n_1,n_2,...,n_M)$ indicates the number of bosons in each orbital, constrained by $\sum_{k=1}^M n_k=N$, with $N$ being the total number of bosons. 
The number of permanents, determined by distributing $N$ bosons across $M$ orbitals, is $ \left(\begin{array}{c} N+M-1 \\ N \end{array}\right)$. 
Here, $|0\rangle$ is the vacuum state, and $\hat{b}_k^\dagger(t)$ represents the time-dependent creation operator for a boson in the $k$-th orbital $\psi_k(x)$, defined as:
\begin{eqnarray}
	\hat{b}_k^\dagger(t)&=&\int \mathrm{d}x \: \psi^*_k(x;t)\hat{\Psi}^\dagger(x;t), \\
	\hat{\Psi}^\dagger(x;t)&=&\sum_{k=1}^M \hat{b}^\dagger_k(t)\psi_k(x;t). \label{eq:def_psi}
\end{eqnarray}
The accuracy of the method is governed by the number of orbitals, $M$. For $M=1$, MCTDH-X reduces to the mean-field Gross-Pitaevskii equation, while for $M \rightarrow \infty$, the wave function becomes exact, as the set $\vert n_1,n_2, \dots ,n_M \rangle$ spans the complete $N$-particle Hilbert space. 
In practice, $M$ is chosen to balance computational feasibility with convergence of observables.

Both the expansion coefficients $C_\mathbf{n}(t)$ and the orbitals $\psi_i(x;t)$ are optimized variationally at every time step~\cite{TDVM81}, allowing for the computation of either ground-state properties through imaginary time propagation or the full time-dependent dynamics through real-time propagation. 
This optimization is performed based on the stationary action principle applied to the many-body action derived from the system's Hamiltonian, expressed in second quantization as:
\begin{align} 
\hat{\mathcal{H}}&=\int dx \hat{\Psi}^\dagger(x) \left\{\frac{p^2}{2m}+V(x)\right\}\hat{\Psi}(x) \nonumber\\
&+\frac{1}{2}\int dx \hat{\Psi}^\dagger(x)\hat{\Psi}^\dagger(x')W(x,x')\hat{\Psi}(x)\hat{\Psi}(x'),
\end{align}
where $V(x)$ represents the one-body potential and $W(x,x')$ describes two-body interactions. 
In this work, $V(x)$ corresponds to the optical lattice potential, while $W(x,x')$ represents dipole-dipole interactions. 
The stationarity of the action leads to a set of coupled equations of motion for the coefficients and orbitals, which are solved simultaneously.

From the optimized orbitals, reduced density matrices can be computed to extract information about system correlations. For instance, the one-body reduced density matrix is given by:
\begin{eqnarray}
\rho^{(1)}(x,x') = \sum_{kq=1}^M \rho_{kq}\psi_k(x)\psi_q(x'),
\label{eq:red-dens-mat}
\end{eqnarray}
where
\begin{eqnarray}
\rho_{kq} = \begin{cases}
\sum_\mathbf{n} |C_\mathbf{n}|^2 n_k, \quad & k=q, \\
\sum_\mathbf{n} C_\mathbf{n}^* C_{\mathbf{n}^k_q} \sqrt{n_k(n_q+1)}, \quad & k\neq q, \\
\end{cases}
\end{eqnarray}
and the summation is over all possible configurations of $\mathbf{n}$. Here, $\mathbf{n}^k_q$ represents the configuration where one boson is removed from orbital $q$ and added to orbital $k$. 
The one-particle density can then be obtained from the diagonal elements of $\rho^{(1)}(x,x')$ as:
\begin{equation}
	\rho(x) = \rho^{(1)}(x,x)/N.
\end{equation}
Higher-order reduced density matrices such as $\rho^{(2)}(x,x')$ etc. can be obtained analogously by tracing over multiple configurations.


\section{System parameters}

In this appendix, we discuss the parameters for the simulations presented in the main text.
Throughout the main text, unless otherwise stated, we perform simulations with $M=12$ orbitals.
The system consists of $N=5$ or $N=8$ bosons in a quasiperiodic optical lattice defined as a superposition of two potentials, a primary lattice of depth $V_p$ and wavelength $\lambda_p$, and a detuning lattice of depth $V_d$, wavelength $\lambda_d$, and phase $\phi$.
This superposition can be written as 
\begin{equation}
V(x) = V_p \sin^2 (k_p x) + V_d \sin^2(k_d x + \phi),
\end{equation}
where $k_i = \frac{2\pi}{\lambda_i}$.
The wavelengths of the two potential are chosen to be compatible with real experimental realizations in ultracold atomic labs, i.e. $\lambda_p \approx 532.2$ nm and $\lambda_d \approx 444.5$ nm, giving an approximately incommensurate ratio $k_d/k_p \approx 1.19721$.
This gives wave vectors $k_p \approx 1.1806 \times 10^7$ m$^{-1}$ and $k_d \approx 1.4135 \times 10^7$ m$^{-1}$. 
For the phase, we choose $\phi = 0.0$ as our results should not depend on the specific location of the disorder.
The potential depths $V_p$ and $V_d$ are tuned across several values to obtain the phase diagram.

\subsection{Lengths}
In our MCTDH-X simulations, we choose to set the unit of length $\bar{L} \equiv \frac{\lambda_p}{2 \pi} = \frac{1}{k_p} = 84.7$ nm, which makes the maxima of the primary lattice appear at multiples of $\pi$ in dimensionless units.
We run simulations with 256 gridpoints in an interval $x \in [-12 \bar{L}, 12 \bar{L}] \approx [-1.016 \mu\mathrm{m}, 1.016 \mu\mathrm{m}]$, giving a resolution of around 3.970 nm.
However, the actual system is defined only for a discrete number of lattice sites in the primary lattice (3 in the commensurate particle filling case and 5 in the incommensurate one).
To restrict the system to the given number of sites, we employ hard-wall boundary conditions at the maxima of the outermost sites.
For example, for 5 sites these hard walls appear at $x = \pm 2.5 \pi \bar{L} \approx \pm 7.85  \bar{L} \approx 665.3$ nm.

\subsection{Energies}
In MCTDH-X, the unit of energy is defined in terms of the unit of length as
$\bar{E} \equiv \frac{\hbar^2}{m} \frac{1}{\bar{L}^2}$.
This can be recast in terms of the recoil energy of the primary lattice, i.e. $E_r \equiv \frac{\hbar^2 k_p^2}{2m} \approx 2.85 \times 10^{-30}$ J with $m \approx 163.93$ Da the mass of $^{164}$Dy atoms.
Since $\bar{L} = \frac{1}{k_p}$, we can immediately see that $\bar{E} = \frac{\hbar^2 k_p^2}{m} = 2 E_r$.

Following recent experiments with quasiperiodic optical lattices, we vary the depths in regimes of several units of the recoil energies.
More specifically,
$V_p \in [3 \bar{E}, 8 \bar{E}] \approx [6.0 E_r,  16 E_r]$, 
$V_d \in [0 \bar{E}, 3 \bar{E}] = [0 E_r,  6.0 E_r]$.
For the dipolar interactions, we probe all regimes from very weak to strongly coupled, i.e. $g_d \in [0.001 \bar{E}, 10.0 \bar{E}] \approx [0.002 E_r, 20.0 E_r]$, which is more than enough to probe all experimentally realizable interaction strengths.

\subsection{Time and ground state relaxations}
Like the unit of energy, the unit of time is defined from the unit of length.
We set $\bar{t} \equiv \frac{m \hat{L}^2}{\hbar} = \frac{m \lambda_p^2}{4 \pi^2 \hbar} = 0.45 \times 10^{-5}$ s.
However, in our simulations, we are only interested in ground state properties and as such we perform evolutions in imaginary time.
We find that 20 units of imaginary time are typically enough to converge to the ground state with an accuracy of 8 decimal digits.

\begin{table}
\centering
\begin{tabular}{ || c | c || }
\hline \hline
Quantity & MCTDH-X units  \\
\hline \hline
unit of length &  $\bar{L} = \frac{1}{k_p} = 84.7$ nm \\
\hline
opt. latt. sites & at $-2 \pi$, $-\pi$, $0$, $\pi$, $2\pi \: [\bar{L}]$ \\
\hline
unit of energy & $\bar{E} = 2 E_r \approx 2.85 \times 10^{-30}$ J \\
\hline
potential depth & $V_p = 8.0 \bar{E} = 16 E_r$ \\
\hline
dipolar interaction & e.g. $g_d=1.0 \bar{E} \approx 2 \: E_r$  \\
\hline
unit of (imaginary) time & $\bar{t} \equiv \frac{m\bar{L}^2}{\hbar} = 0.45 \times 10^{-5}$ s. \\
\hline \hline
\end{tabular}
\caption{Units used in MCTDH-X simulations. $E_r=\frac{\hbar^2 k_p^2}{2m}$ is the recoil energy of the primary lattice.}
\end{table}

\section{Density at very weak interactions}
\label{app:ultraweak}
\begin{figure}[h!]
\centering
\includegraphics[width=1.0\columnwidth]{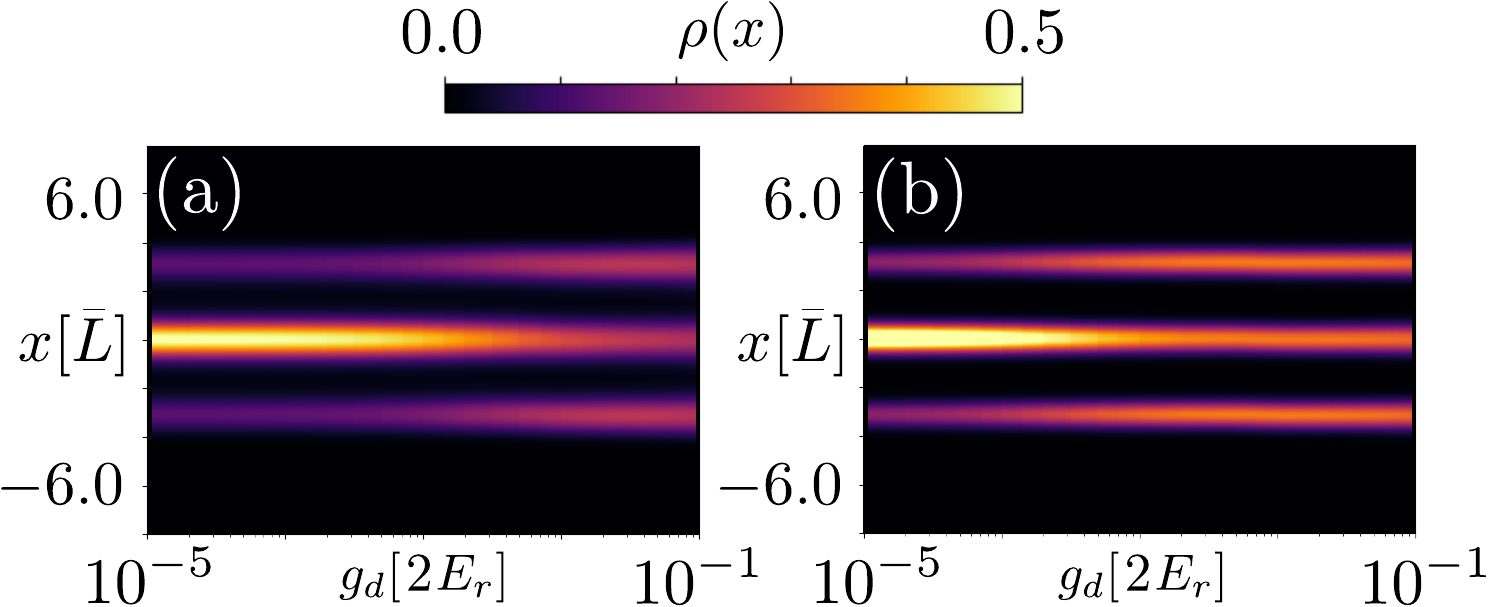}
\caption{Real-space density distribution for the ground state of $N=6$ bosons in $S=3$ sites (commensurate filling) with no disorder, plotted as a function of dipolar interaction strength $g_d$ in the ultraweak limit.
The two panels refer to different values of the primary optical lattice:
(a) $V_p = 6.0 E_r$ and (b) $V_p = 16.0 E_r$.
}
\label{fig:density-ultraweak}
\end{figure}

In this appendix, we provide an additional figure that visualizes the density profile in the periodic case at ultraweak interactions, demonstrating that the ground state in this case tends towards a superfluid confined mostly to a single site.
This is shown in Fig.~\ref{fig:density-ultraweak}

%
%
%

\bibliography{biblio}

\end{document}